\documentclass[twocolumn,amssymb,superscriptaddress,aps,prd,floatfix]{revtex4}
\usepackage{amsmath}
\usepackage[dvips]{graphicx}
\usepackage[dvips]{color}
\usepackage[usenames,dvipsnames]{xcolor}
\newlength{\colw}
\newcommand{\err}[2]{\mbox{$\stackrel{\scriptstyle +#1}{\scriptstyle -#2}$}}

\newcommand{\bra}{\langle}
\newcommand{\ket}{\rangle}
\newcommand{\braket}[1]{\langle#1\rangle}

\newcommand{\order}{{\cal O}}
\newcommand{\One}{1\kern-4.5pt1}

\newcommand{\qq}{\bra qq\ket}

\newcommand{\latt}{{\text{lat}}}
\newcommand{\cont}{{\text{cont}}}

\newcommand{\del}[2]{\frac{\partial#1}{\partial#2}}

\newcommand{\cdeconf}{\cite{Hands:2006ve}}
\newcommand{\cquarkyonic}{\cite{Hands:2010gd}}
\newcommand{\cprev}{\cite{Hands:2010gd,Cotter:2012mb,Boz:2013rca}}
\newcommand{\callprev}{\cite{Hands:2006ve,Hands:2010gd,Cotter:2012mb,Boz:2013rca}}
\newcommand{\cphases}{\cite{Cotter:2012mb}}
\newcommand{\ctrans}{\cite{Boz:2013rca}}

\begin{document}

\title{Dense two-color QCD towards continuum and chiral limits}

\author{Tamer Boz}

\affiliation{
Department of Theoretical Physics, National University of Ireland Maynooth,
Maynooth, County Kildare, Ireland.}

\author{Pietro Giudice}

\affiliation{
Istituto Arcivescovile Paritario Santa Caterina, 
Piazza Santa Caterina 4, 56127 Pisa, Italy
}
\affiliation{
       INFN Sezione di Pisa, Largo Pontecorvo 3, 56127 Pisa, Italy.
}

\author{Simon Hands}

\affiliation{
       Department of Physics, College of Science, Swansea University,
       Singleton Park, Swansea SA2 8PP, U.K.
       }

\author{Jon-Ivar Skullerud\thanks{
On leave at
Galileo Galilei Institute for Theoretical Physics, 
Largo Enrico Fermi 2, 50125 Firenze, Italy.}
}

\affiliation{
Department of Theoretical Physics, National University of Ireland Maynooth,
Maynooth, County Kildare, Ireland.}
\affiliation{
School of Mathematics, Trinity College, Dublin 2, Ireland}

\begin{abstract}
We study two-color QCD with two flavors of Wilson fermion as a
function of quark chemical potential $\mu$ and temperature $T$, for
two different lattice spacings and two different quark masses. We find
that the quarkyonic region, where the behaviour of the quark number
density and the diquark condensate are described by a Fermi sphere of
almost free quarks distorted by a BCS gap, extends to larger chemical
potentials with decreasing lattice spacing or quark mass.  In both
cases, the quark number density also approaches its non-interacting
value.  The pressure at low temperature is found to approach the
Stefan--Boltzmann limit from below.
\end{abstract}

         
\maketitle

\section{Introduction}

The structure of strongly interacting matter at high densities and low
to moderate temperatures remains an outstanding problem, with
applications to compact stars, neutron star mergers, and the next
generation of heavy-ion colliders at FAIR and NICA.  First-principles
studies of this regime are hindered by the sign problem: with chemical
potential $\mu\neq0$ the euclidean
action becomes complex, and can therefore not be used as a probability
weight in Monte Carlo simulations, which are the mainstay of lattice
gauge theory, the method of choice for first-principles,
non-perturbative quantum field theory.  Despite recent progress in
alternative sampling approaches such as the density of states method
\cite{Langfeld:2016kty}, complex Langevin \cite{Seiler:2017wvd} and
Lefschetz thimble and related approaches
\cite{Scorzato:2015qts,Alexandru:2015sua}, we do not as yet have any
method that has been shown to yield valid and reliable results for
real QCD.

The problem may be circumvented by studying QCD-like theories 
without a sign problem, such as theories with adjoint fermions in any
gauge group, QCD with isospin chemical potential
\cite{Brandt:2017oyy}, or QCD with gauge groups SU(2) (QC$_2$D)
\cprev\ or G$_2$ (G$_2$-QCD) \cite{Maas:2012wr,Wellegehausen:2013cya}.  Although these
theories all have qualitative features which distinguish them from
real QCD at nonzero baryon chemical potential --- notably a
gauge-invariant Bose--Einstein condensate (BEC) above an onset
chemical potential $\mu_o$ --- they share salient features such as
spontaneous chiral symmetry breaking and confinement at $T=\mu=0$, and may be used
as laboratories for strongly interacting theories at high density.
Lattice results from these theories may also be used as a check on the
approximations made in other approaches which do not suffer from the
sign problem, including Polyakov-loop extended Nambu--Jona-Lasinio
models \cite{He:2010nb,Andersen:2015sma}, massive perturbation theory \cite{Kojo:2014vja,Suenaga:2019jjv}, quark--meson(--diquark) coupling
models \cite{Strodthoff:2013cua}, the functional renormalisation group
\cite{Khan:2015puu} or Dyson--Schwinger equations
\cite{Contant:2017gtz,Contant:2019lwf}.

In a previous series of papers \callprev, we have studied the phase
structure of QC$_2$D with $N_f=2$ Wilson fermions.  The main findings
of these studies have been that at high density and low temperature,
there is a `quarkyonic' phase \cquarkyonic\ where the diquark
condensate and quark number density scale with the quark chemical
potential $\mu$ in the same way as in a system composed of
noninteracting fermions disrupted by a Bardeen--Cooper--Schrieffer
(BCS) gap.  The diquark condensate, signalling superfluidity, vanishes
at a critical temperature which appears to be approximately
independent of $\mu$ above the onset chemical potential
$\mu_o=m_\pi/2$ \ctrans.  At high temperature, there is a transition
to a deconfined quark--gluon plasma, with the pseudocritical
temperature $T_d$ decreasing with increasing $\mu$ \ctrans.  It is as
yet unclear whether $T_d$ goes to zero at any finite $\mu$ and hence
whether there is any deconfinement transition at high density and low
temperature; early indications of such a transition \cdeconf\ may have
been complicated by lattice artefacts.

These studies have all been carried out with quite heavy quarks
($m_\pi/m_\rho=0.8$) and on fairly coarse lattices
($a=0.18-0.23\,$fm).  The aim of the current paper is firstly to gain
control over lattice artefacts by reducing the lattice spacing at
fixed $m_\pi/m_\rho$, and secondly to explore the quark mass
dependence by studying a system with lighter quarks at fixed lattice
spacing.  The latter is of particular significance as it might point
to a ``BEC region'' which can be described using chiral perturbation
theory ($\chi$PT) incorporating both mesonic and baryonic (diquark) Goldstone
degrees of freedom \cite{Kogut:2000ek}.

There have been a number of other lattice studies of dense QC$_2$D in
recent years, using staggered
\cite{Braguta:2016cpw,Astrakhantsev:2018uzd,Wilhelm:2019fvp} and
Wilson \cite{Iida:2019rah} fermions.  These have to a large extent
confirmed the picture outlined above, with some additions.  Notably,
in \cite{Braguta:2016cpw}, with a smaller pion mass than in \cprev, a
BEC region was found where the diquark condensate and quark number
density agree with predictions from chiral perturbation theory,
followed by a transition to a quarkyonic region at higher $\mu$.
Also, the chiral condensate was found to vanish in the chiral limit in
both the BEC and quarkyonic regions.  The system was found to be
confined at low temperature, with deconfinement only setting in at
much larger chemical potential \cite{Astrakhantsev:2018uzd}.  Similar
conclusions were found in \cite{Iida:2019rah}.

It is also worth noting that a recent study of QCD with nonzero
isospin chemical potential \cite{Brandt:2017oyy} found a phase diagram
very similar to that of \ctrans, namely a pion condensed
phase at low $T$ and large $\mu$, with a critical temperature that is
nearly independent of $\mu$, and a deconfinement transition line that
intersects with the pion condensation transition.

The structure of this paper is as follows.  In
section~\ref{sec:simulation} we describe our simulation parameters and
determination of the lattice spacing for our new (fine) ensemble.
Results from this ensemble are presented in
section~\ref{sec:results-fine}.  Firstly, in
sec.~\ref{sec:diquark-fine} we present results for the superfluid
order parameter, the diquark condensate, including its scaling with
chemical potential and estimates for the critical temperature.
Sec.~\ref{sec:deconf-fine} contains our results for the Polyakov loop and
deconfinement transition, while sec.~\ref{sec:nq-fine} contains results for the
quark number density.  Section~\ref{sec:results-light} contains our
results from simulations with lighter quarks.  We summarise our
findings in section~\ref{sec:conclude}.

\section{Simulation details and scale setting}
\label{sec:simulation}

We study QC$_2$D with a conventional Wilson action for the gauge
fields and two flavours of Wilson fermion.  The fermion action is
augmented by a gauge- and iso-singlet diquark source term which serves
the dual purpose of lifting the low-lying eigenvalues of the Dirac
operator and allowing a controlled study of diquark condensation.  The
quark action is
\begin{equation}
S_Q+S_J=\sum_{i=1,2}\bar\psi_iM\psi_i
 + \kappa j[\psi_2^{tr}(C\gamma_5)\tau_2\psi_1-h.c.],
\label{eq:Slatt}
\end{equation}
where $i=1,2$ is a flavour index and
\begin{align}
M_{xy}=\delta_{xy}-\kappa\sum_\nu&\Bigl[(1-\gamma_\nu)e^{\mu\delta_{\nu0}}U_\nu(x)\delta_{y,x+\hat\nu}\nonumber\\
&+(1+\gamma_\nu)e^{-\mu\delta_{\nu0}}U^\dagger_\nu(y)\delta_{y,x-\hat\nu}\Bigr].
\label{eq:Mwils}
\end{align}
Further details about the action and the Hybrid Monte Carlo algorithm
used can be found in \cite{Hands:2006ve}.

\begin{table}
\begin{tabular*}{\colw}{l@{\extracolsep{\fill}}llllll}
\hline
Name & $\beta$ & $\kappa$& $am_\pi$&
$m_\pi/m_\rho$ & $a$ (fm) \\ \hline
Light &1.7 & 0.1810 & 0.438(15) & 0.61(5) &  0.189(4) \\
Coarse & 1.9 & 0.1680 & 0.645(8) & 0.805(9) & 0.178(6) \\
Fine & 2.1 & 0.1577 & 0.446(3) & 0.810(7) & 0.138(6) \\
\end{tabular*}
\caption{Simulation parameters, pion and rho meson masses and
  lattice spacing at $\mu=j=0$.}
\label{tab:params-mu0}
\end{table}

\begin{figure}[tbp]
\begin{center}
\includegraphics*[width=\colw]{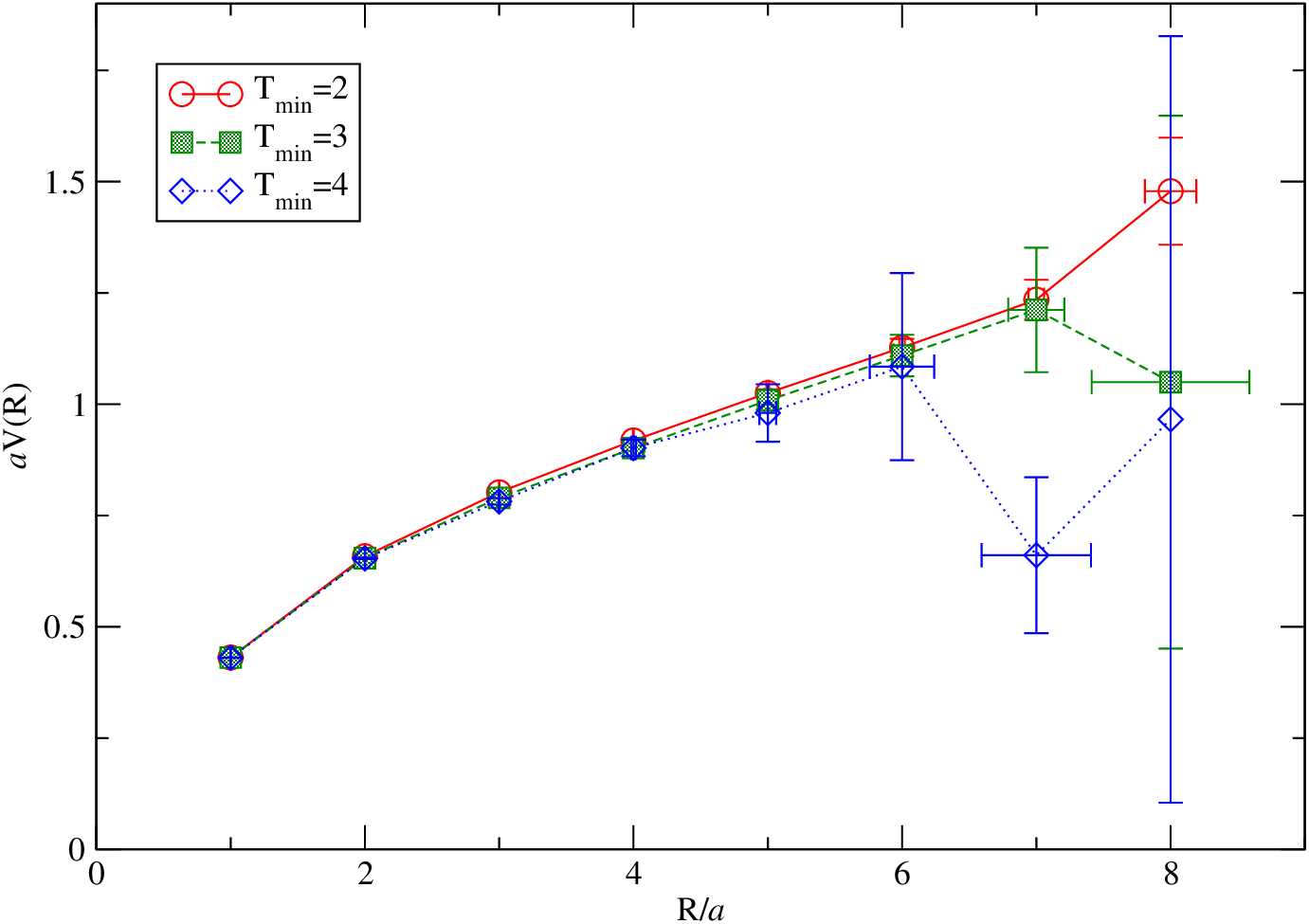}
\caption{Static quark potential versus spatial separation of the
  Wilson loop on the fine ensemble for $T=\mu=0$, for different values of the
  minimum time extent $T_{\min}$ of the Wilson loop.}
\label{fig:static-potential}
\end{center}
\end{figure}

We have studied three ensembles, which in the following we call
`coarse', `fine' and `light'.  The parameters are shown in
Table~\ref{tab:params-mu0}, together with the values obtained for the
pion (pseudoscalar meson) mass $m_\pi$, ratio of pion to rho (vector
meson) mass $m_\pi/m_\rho$ and lattice spacing $a$.  The coarse
ensemble is the same as was used in \cprev.  The parameters for the
fine ensemble were chosen to give the same value of $m_\pi/m_\rho$ as
the coarse ensemble, while those of the light ensemble were chosen to
give approximately the same lattice spacing as the coarse ensemble,
but with a smaller value of $m_\pi/m_\rho\approx0.6$.  Further details
about the coarse and light ensemble parameters can be found in
\cphases.

To determine the lattice spacing, we extracted the static quark
potential $V(r)$ from rectangular Wilson loops $W(r,\tau)$ by fitting
$W(r,\tau)=\exp(-V(r)\tau)$ for $\tau/a=T_{\min},N_\tau-1$.
The lattice
spacing was then determined by fitting the static quark potential to the
Cornell form 
\begin{equation}
V(r)=C+\alpha/r+\sigma r\,,
\label{eq:Cornell}
\end{equation}
and taking the string tension
to be $\sqrt{\sigma}=440$MeV.  The static quark potential for the
fine ensemble is shown in Fig.~\ref{fig:static-potential} and the
best fit values are given in Table~\ref{tab:scale-set}. 

\begin{table}[tb]
\begin{center}
\begin{tabular}{|c|ccccc|}
\hline
$T_{\min}$ & $\sigma a^2$ & $\alpha$ & $C a$ & $\chi^{2}/N_{dof}$ & $a$ (fm) \\
\hline
2 & $0.1283\err{9}{9}$ & $0.2179\err{21}{20}$ & $0.5224\err{29}{29}$ & 2.45 & $0.1706\err{6}{6}$\\ 
3 &  $0.0987\err{15}{15}$ & $0.2578\err{36}{32}$ & $0.5902\err{50}{51}$ & 1.25 & $0.1496\err{12}{12}$\\ 
4 & $0.0840\err{65}{86}$ & $0.2779\err{146}{206}$ & $0.6245\err{286}{200}$ & 0.249 & $0.1380\err{52}{72}$\\ 
\hline 
\end{tabular}
\end{center}
\caption{Fit parameters for fits of the static quark potential on the
  fine ensemble to the
  Cornell form \eqref{eq:Cornell}, and the corresponding lattice
  spacing values found for various minimum temporal extents $T_{\min}$
  of the Wilson loop.}
\label{tab:scale-set}
\end{table}

\begin{table}[tb]
\begin{tabular}{c|rrccc}
Ensemble & $N_s$ & $N_\tau$ & $T$ (MeV) &  $\mu a$ & $\mu$ (MeV) \\ \hline
Fine & 16 & 32 & 45 & 0.1--0.8 & 143--1142 \\
     & 16 & 20 & 71 & 0.1--0.7 & 143--999 \\
     & 16 & 16 & 89 & 0.1--0.6 & 143--857 \\
     & 16 & 12 & 119 & 0.1--0.6 & 143-857 \\\hline
Coarse & 12 & 24 & 47 & 0.25--1.1 & 277--1217 \\
       & 16 & 12 & 94 & 0.4--0.9 & 443--996 \\
       & 16 & 8 & 141 & 0.4--0.7 & 443--775 \\ \hline
Light  & 12 & 24 & 43 & 0.1--0.8 & 104--834 \\
       & 16 & 12 & 87 & 0.1--0.8 & 104--834 \\
       & 16 & 8 & 130 & 0.1--0.7 & 104--730 \\ \hline
\end{tabular}
\caption{Temperatures $T$ and chemical potential values
$\mu$  used in this study.}
\label{tab:T-mu-params}
\end{table}

We have performed scans in $\mu$ at fixed temperature for four
temperatures on the fine ensemble, and three 
temperatures on the light ensemble.  The lattice volumes, temperatures
and range of chemical potentials are given in
table~\ref{tab:T-mu-params}, along with those from the coarse ensemble
that are used for comparison.  For most of these temperatures and
chemical potentials we have produced configurations at two values for
the diquark source, namely $ja=0.02, 0.03$ on the fine ensemble, and
$ja=0.02, 0.04$ on the light ensemble.  A third diquark source value
($ja=0.01$ for the fine ensemble, $ja=0.03$ for the light ensemble)
has been added for selected parameter values to control the
systematics of the $j\to0$ extrapolations.  For a few parameters,
mostly at the highest and lowest $\mu$-values, only a single $j$-value
has been used; in these cases, only results for the Polyakov loop, which
is only weakly dependent on $j$, will be shown here.  In addition, we
have performed temperature scans at fixed values of
$\mu a=0,0.2,0.3,0.4,0.5$ on the fine ensemble, with $N_\tau=18$--4
corresponding to $T=$63--356\,MeV.

We note that by assigning a temperature $T>0$ to our ensembles with
$N_\tau\geq N_s$ we deviate from what is commonly done in lattice
thermodynamics studies.  This is appropriate in the presence of a
chemical potential (which modifies the temporal boundary
conditions). At weak coupling and low temperature, relativistic quarks 
form a Fermi surface with Fermi momentum $k_F\simeq\mu_q$.  The
resulting ground state is highly degenerate, so that unlike the case
when $\mu=0$ the discrete momenta on the finite volume do not lead to
a large energy gap.  For this reason in our view one should
always consider $T$ as non-zero when studying systems with $\mu_q\neq0$,
even if $N_s<N_\tau$.

For enhanced separation of scales $m_\pi\ll m_\rho$, the behaviour as $\mu$
increases at zero temperature may be analysed using $\chi$PT~\cite{Kogut:2000ek},
in this context an effective theory of tightly-bound $q\bar q$ mesons
and $qq, \bar{q}\bar{q}$
baryons. For $\mu=0$ chiral symmetry of continuum QC$_2$D is spontaneously broken from SU($2N_f$)
to Sp($2N_f$) yielding $N_f(2N_f-1)-1$ Goldstones; for $N_f=2$ these are the
pseudoscalar pion triplet and a scalar diquark/antidiquark pair. As $\mu$ is raised there is a second-order
\emph{onset} transition at $\mu_o=m_\pi/2$ to a phase where the baryon density $n_q$
and superfluid diquark condensate $\qq$ are both non-zero:
\begin{align}
\frac{\qq}{\braket{\bar{q}q}_0}
 &=\sqrt{1-\frac{\mu_o^4}{\mu^4}}\theta(\mu-\mu_o)\label{eq:qq-chiPT}\,,\\
n_q&= 8N_fF_\pi^2\mu\left(1-\frac{\mu_o^4}{\mu^4}\right)\theta(\mu-\mu_o)\,.\label{eq:nq-chiPT}
\end{align}
Here $\braket{\bar{q}q}_0$ is the chiral condensate at $\mu=0$ and $F_\pi$ 
the $\chi$PT parameter known as the pion decay constant. 
Note that the onset transition in physical QCD is first order.

\section{Results from fine ensemble}
\label{sec:results-fine}

\subsection{Diquark condensation}
\label{sec:diquark-fine}

\begin{figure}[tb]
\includegraphics*[width=\colw]{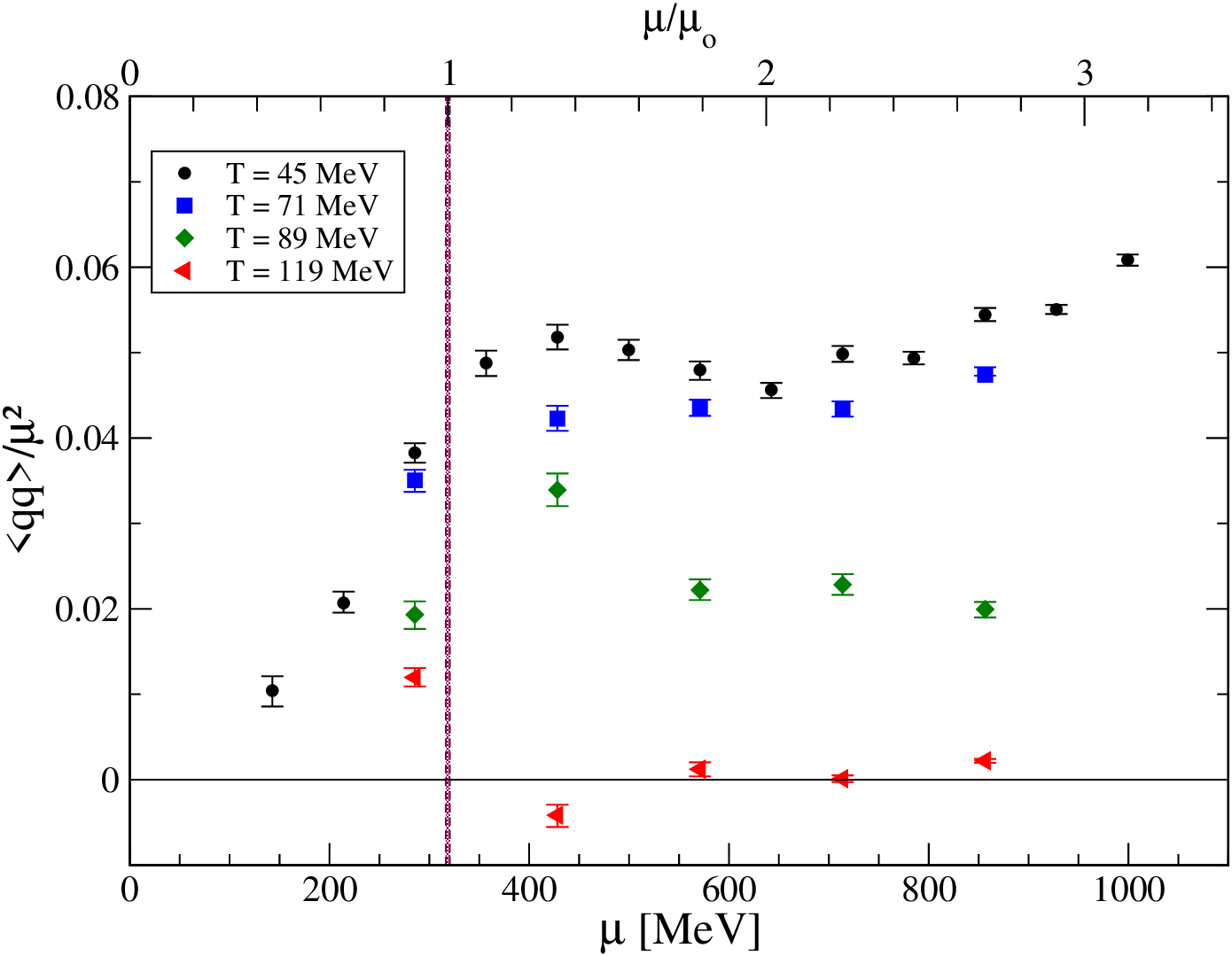}
\caption{The diquark condensate $\bra qq\ket/\mu^2$
on the fine ensemble, extrapolated to $j=0$ for $N_\tau=32, 20, 16, 12$
($T=45,71,89,119$ MeV). The vertical dashed lines denote the onset
transition at zero temperature, $\mu_o=m_\pi/2$.}
\label{fig:qq-fine}
\end{figure}

Figure~\ref{fig:qq-fine} shows the diquark condensate,
\begin{equation}
\qq \equiv \del{\ln{\cal Z}}{j}
 = \frac{\kappa}{2}\bra\psi^{2tr}C\gamma_5\tau_2\psi^1
  -\bar\psi^1C\gamma_5\tau_2\bar\psi^{2tr}\ket\,,
\label{def:qq}
\end{equation}
divided by the
square of the chemical potential,
as a function of chemical potential, for all temperatures.  
Physically, $\qq$ is the density of relativistic quark pairs
contributing to the superfluid condensate.
In the case of a weakly-coupled BCS condensate at the Fermi
surface, this should be roughly equal to the momentum-space volume
of a shell centred on the Fermi surface with thickness
$\order(\Delta)$, the superfluid gap; by contrast leading order 
$\chi$PT \eqref{eq:qq-chiPT}  predicts $\bra
qq\ket$ to be $\mu$-independent for $\mu\gg\mu_o$ and hence the
quantity plotted should fall off like $\mu^2$.

For the two lowest temperatures, we find that $\bra qq\ket/\mu^2$ is
almost independent of both $\mu$ and $T$ for $\mu\gtrsim\mu_o$.  This
agrees with what was found previously for the coarse ensemble \cphases.  For
$\mu<\mu_o$ the diquark condensate rises gradually from zero; we take
this to be primarily an artefact of the linear extrapolation in $j$.

\begin{figure}[tb]
\includegraphics*[width=\colw]{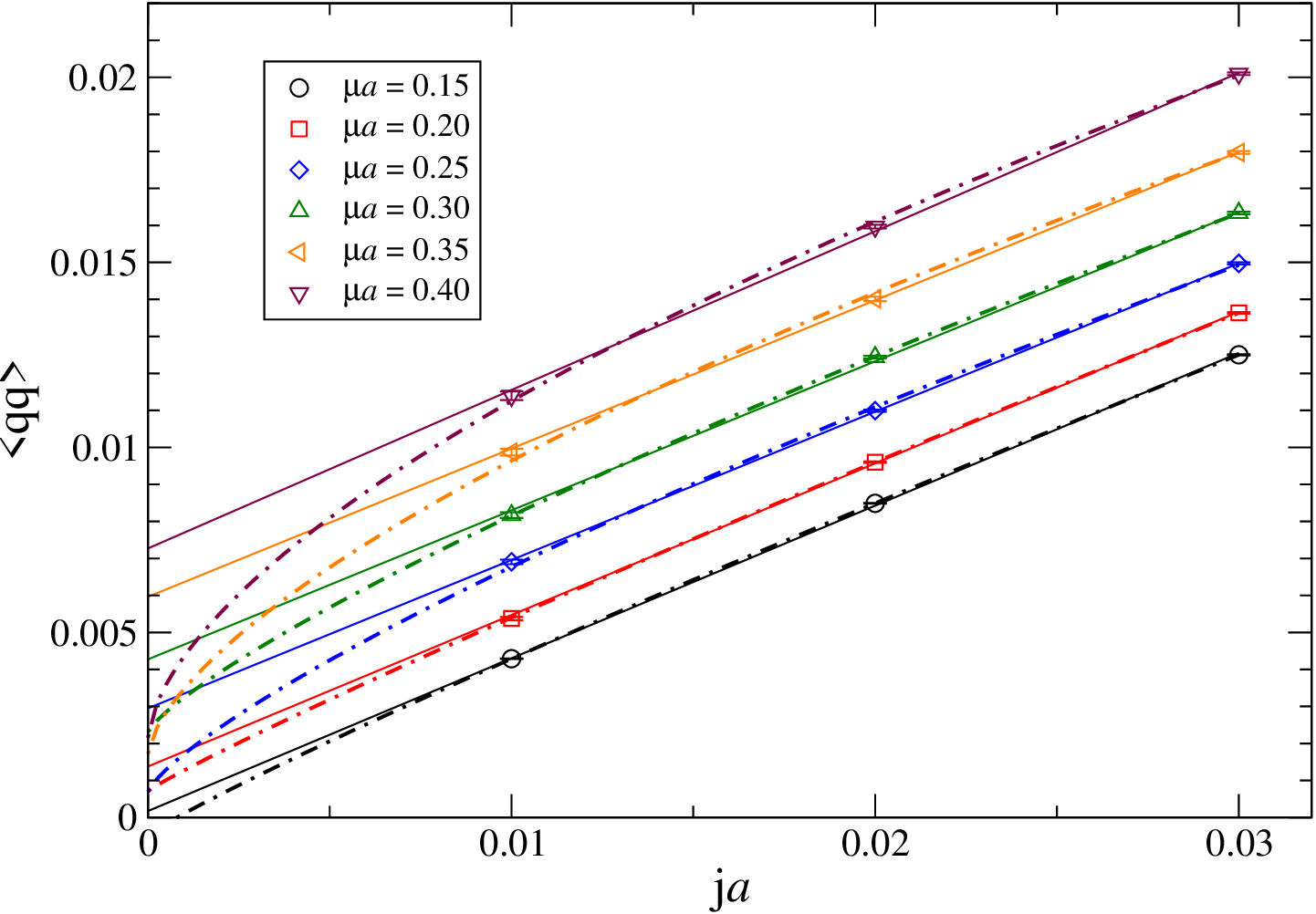}
\caption{The diquark condensate $\qq$
on the fine ensemble, as a function of the diquark source, for
different values of the chemical potential $\mu$.  The solid lines
are linear fits to the data, while the dash-dotted lines are fits to a
power-law + constant form $\qq=A+Bj^\alpha$.}
\label{fig:qq-fine-jextrap}
\end{figure}

To further investigate the diquark source dependence, in
fig.~\ref{fig:qq-fine-jextrap} we show $\qq$ as function of
$j$ for different values of the chemical potential.  With our current
data, we see no evidence of deviation from a linear form at any $\mu$.
Specifically, the form $\qq=Aj^{1/3}$, which should hold exactly at
$\mu=\mu_o$, does not fit the data for any of our
$\mu$-values, even with the addition of a constant term.  Fits to a
more general power-law form, $\qq=A+Bj^\alpha$, yield powers ranging
from $\alpha\sim0.9$ for $\mu<\mu_o$ to $\alpha\sim0.6$ at $\mu
a=0.4$.  The apparent linear $j$-dependence at all $\mu$ may be an
indication that non-analytic behaviour only sets in at lower
$j$-values than we have here.  To improve on the diquark source
extrapolation, it may be necessary to carry out a reweighting
procedure as outlined in \cite{Brandt:2017oyy}, or explore much
smaller values of the source $ja$~\cite{Braguta:2016cpw}.

At $T=90$ MeV ($N_\tau=16$) we see that $\qq$ is significantly smaller
for all values of $\mu$, suggesting that this temperature is near the
critical temperature for the superfluid to normal phase transition.
This is again in agreement with what was found for the coarse lattice 
in \cite{Cotter:2012mb,Boz:2013rca}.  Finally, at the highest
temperature, $T=120\,$MeV ($N_\tau=12$), the diquark condensate is
consistent with zero, suggesting that at this temperature we are in
the normal phase.

\begin{figure}[tb]
\includegraphics*[width=\colw]{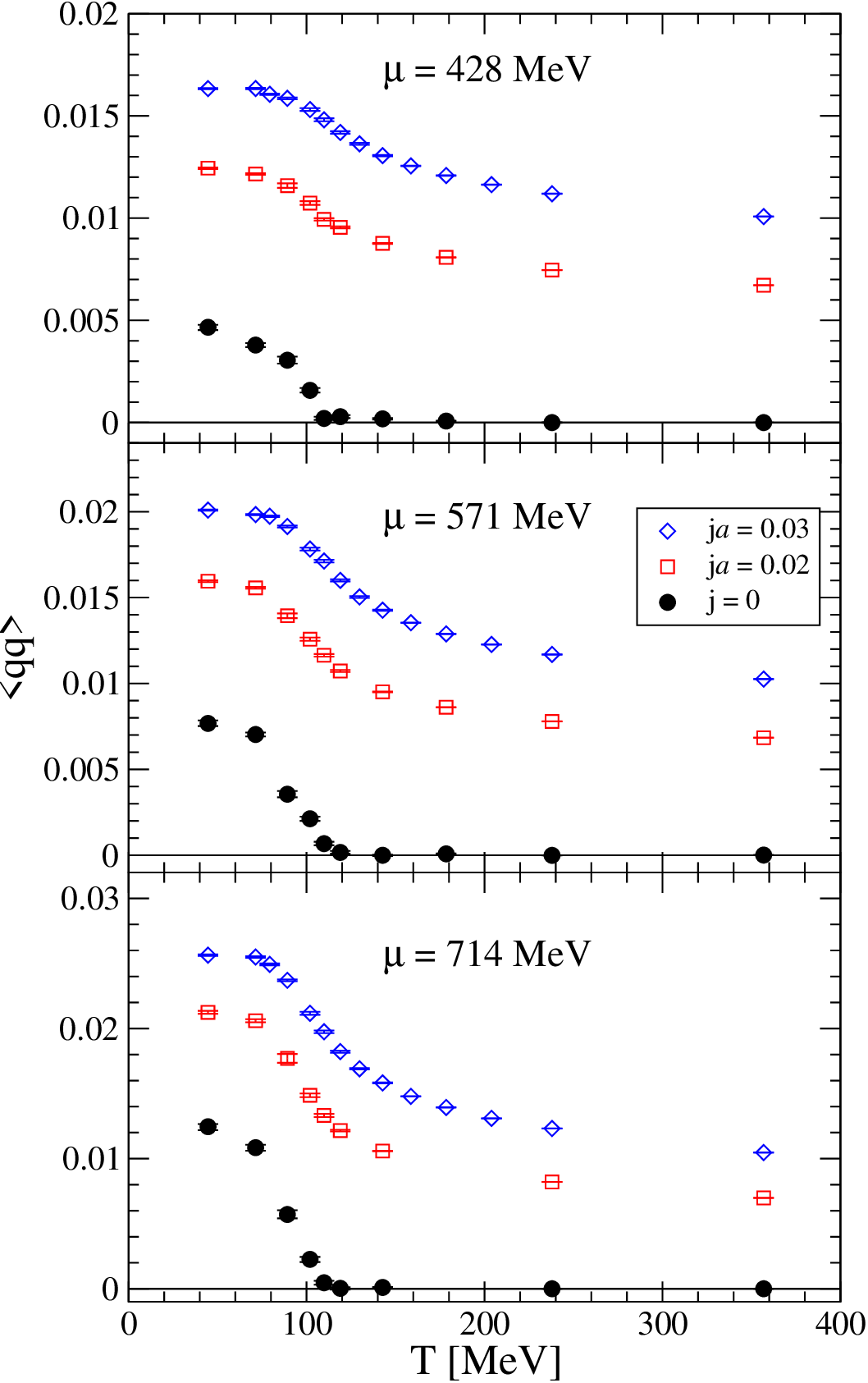}
\caption{The diquark condensate $\bra qq\ket$
on the fine ensemble, as a function of temperature, for chemical
potentials $\mu=428, 571, 714\,\mathrm{MeV} (a\mu=0.3, 0.4, 0.5)$.}
\label{fig:qq-fine-Tscan}
\end{figure}

To study the superfluid to normal phase transition in more detail, we
have performed temperature scans at fixed values of chemical potential
$a\mu=0.3, 0.4, 0.5$ and $aj=0.02, 0.03$.  These are fixed-scale
temperature scans; i.e., the temperature is varied by changing
$N_\tau$, without changing the lattice spacing.  This means that
although our data for $\qq$ are not renormalised, this will just
contribute an overall factor without changing the shape of the curves.

The results of these
scans are shown in fig.~\ref{fig:qq-fine-Tscan}.  From the data
extrapolated to $j=0$ using a linear Ansatz, we see evidence of a
phase transition at $T_s\approx110\,$MeV,
independent of the chemical potential.  In order to determine this
transition in a more controlled manner, we find the inflection
points for $ja=0.03$ and 0.02 using a cubic spline interpolation, and
extrapolate these to $j=0$.  The results are shown in table~\ref{tab:Ts}.
This yields a somewhat lower temperature 
$T_s=$90--100\,MeV, which is consistent with the result quoted in
\ctrans, $T_s=93(8)\,$MeV.  We note that the data are for
a single volume and we can therefore not determine the order of the
transition, but it is expected to be a second order transition (in the
O(2) universality class), and the data are consistent with this.

\begin{table}[tb]
\begin{tabular}{r|ccc}
    $a\mu$ & 0.3 & 0.4 & 0.5 \\\hline
$aT_s(0.03)$ & 0.0782\err{14}{26}\err{3}{1} 
 & 0.0711\err{43}{12}\err{86}{0} & 0.0715\err{6}{5}\err{0}{33} \\
$aT_s(0.02)$ & 0.0734\err{2}{3}\err{7}{8}
 & 0.0722\err{7}{17}\err{18}{0} & 0.0677\err{38}{11} \\ \hline
$aT_s$ & 0.063(6)(3) & 0.075(8)(10) & 0.060(8)(6) \\
$T_s$ (MeV) & 90(10) & 107(18) & 86(10) 
\end{tabular}
\caption{Inflection points $T_s(j)$ for $\qq(T)$ at $ja=0.03,0.02$
    and critical temperature $T_s$ obtained from extrapolating
    $T_s(j)$ to $j=0$.  The first set of uncertainties is
    statistical; the second set is due to systematic uncertainties in the
    interpolation.}
  \label{tab:Ts}
\end{table}

\subsection{Deconfinement}
\label{sec:deconf-fine}

\begin{figure}[tb]
\includegraphics*[width=\colw]{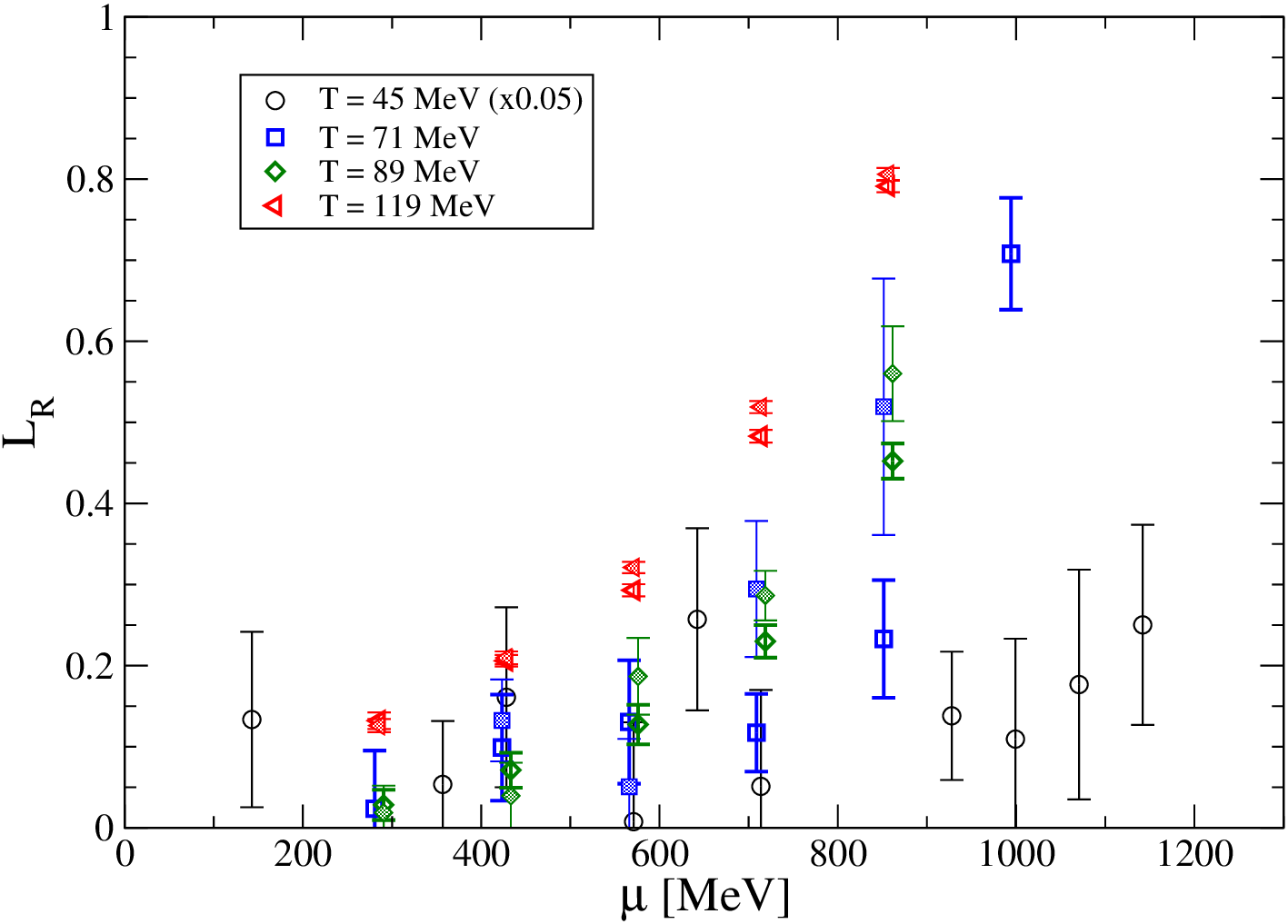}
\caption{The renormalised Polyakov loop using Scheme A, as a function
  of chemical potential, for all temperatures.  The open symbols are
  for $ja=0.03$; the shaded symbols are for $ja=0.02$.
}
\label{fig:polyakov-fine-mu}
\end{figure}

\begin{figure}[tb]
\includegraphics*[width=\colw]{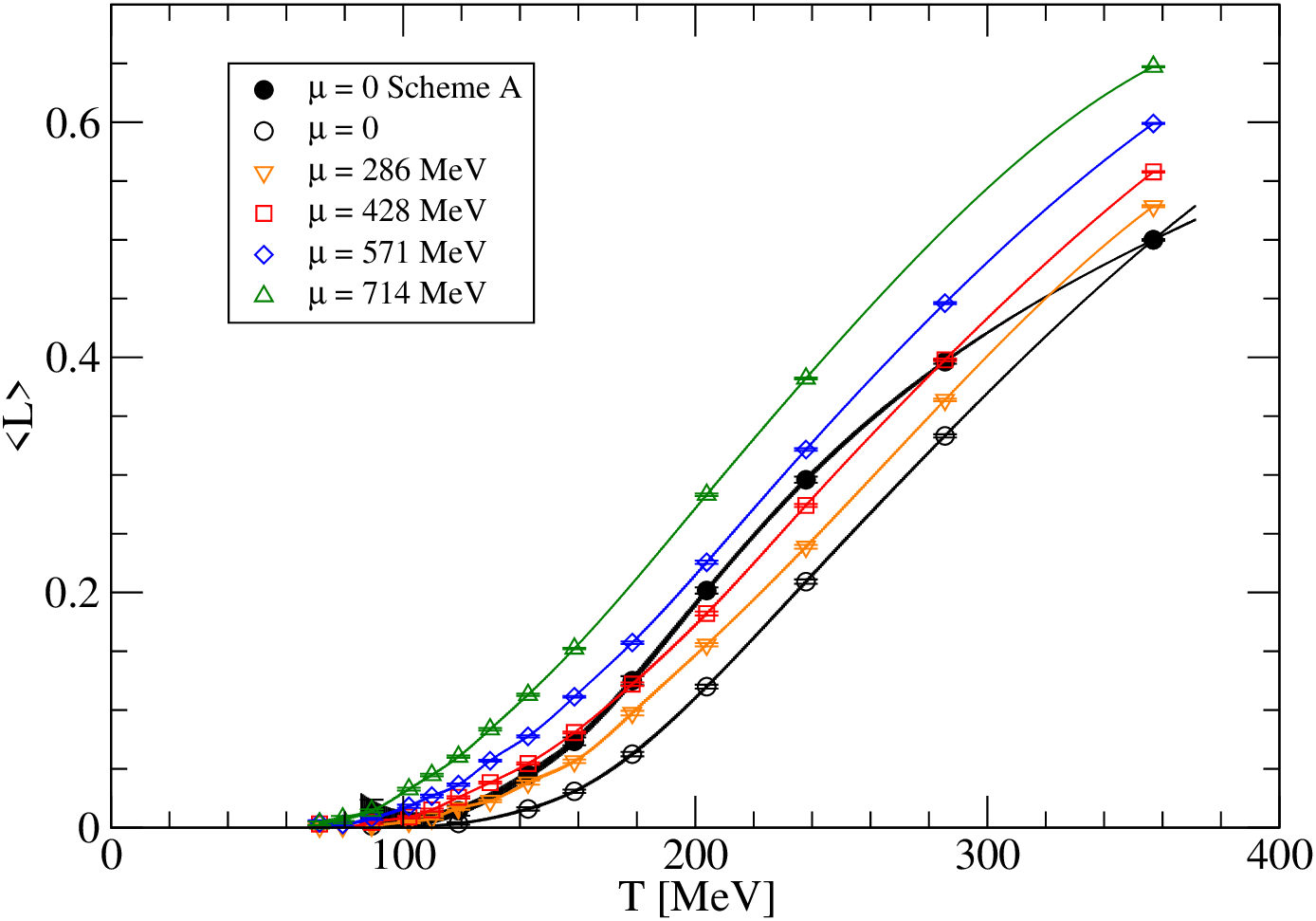}
\caption{The renormalised Polyakov loop $\braket{L}$ in Scheme B as a
  function of temperature $T$ for $ja=0.03$ and  chemical potentials
  $\mu$ corresponding to $\mu a=0.2,
  0.3,0.4,0.5$.  Also shown in the $\mu=0$ Polyakov loop renormalised
  according to Scheme A, divided by 2 for ease of comparison.}
\label{fig:polyakov-fine-Tscan}
\end{figure}
\begin{figure}[tb]
\includegraphics*[width=\colw]{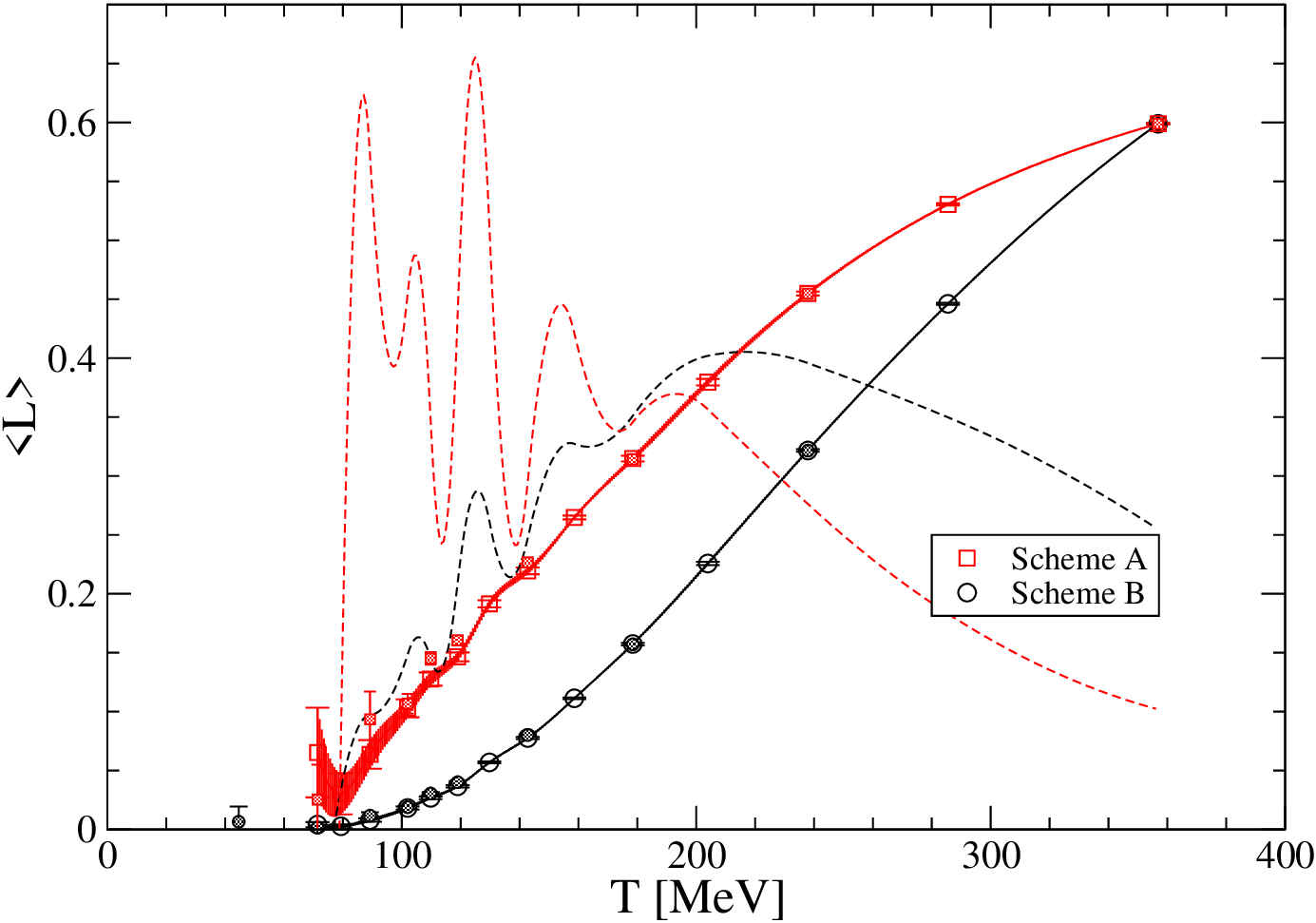}
\caption{The renormalised Polyakov loop $\braket{L}$ as a function of
  temperature $T$ for $ja=0.03$ and $\mu a=0.4$, with
  two different renormalisation schemes: Scheme A (red squares) and
  Scheme B (black circles), see text for details.  Scheme A data have
  been divided by 2 to ease the comparison between the schemes. The
  solid (dashed) lines are the derivatives of cubic spline
  interpolations of the data 
  points for Scheme A (B).  The smaller, shaded symbols are results
  for $ja=0.02$.}
\label{fig:polyakov-fine-schemeAB}
\end{figure}

Figure~\ref{fig:polyakov-fine-mu} shows the order parameter for deconfinement,
the Polyakov loop $\bra L\ket$, for our four different temperatures.
The renormalised Polyakov loop $L_R$ is given in terms of the bare
Polyakov loop $L_0$ and the temperature $T=1/(aN_\tau)$ by
\begin{equation}
L_R(T,\mu)= Z_L^{N_\tau}L_0(\frac{1}{aN_\tau},\mu)\,.
\label{eq:polyakov-renorm}
\end{equation}
Just as in \ctrans, we use two different renormalisation schemes,
\begin{align*}
\textbf{Scheme A} & \qquad & L_R(T=\frac{1}{4a},\mu=0) &= 1\,,\\
\textbf{Scheme B} & \qquad & L_R(T=\frac{1}{4a},\mu=0) &= 0.5\,.
\end{align*}
The results in fig.~\ref{fig:polyakov-fine-mu} have been obtained
using Scheme A.  We see no evidence of any deconfinement transition at
our lowest temperature, $T=45\,$MeV, corresponding to $N_\tau=32$.
This suggests that the 
high-density deconfinement transition found on coarser lattices at
temperatures of 40--50\,MeV \cite{Hands:2006ve,Cotter:2012mb} is
primarily a lattice artefact.

At higher temperatures we see that $\bra L\ket$ increases rapidly from
zero above a chemical potential $\mu_d(T)$ which is a decreasing
function of temperature, in agreement with previous results.

To determine the transition line, we study the variation of the
renormalised Polyakov loop with temperature at fixed values of the
chemical potential $\mu a=0.2, 0.3, 0.4, 0.5$.  The results from
Scheme B are shown
in fig.~\ref{fig:polyakov-fine-Tscan}.  The results are indicative of
a broadening of the transition and a reduction in the transition
temperature as the chemical potential is increased.  However, this
effect is not large: the inflection point in Scheme B actually appears
to increase between $\mu=0$ and $\mu a=0.2$, and even at our highest
chemical potential, $\mu a=0.5$, the inflection point in Scheme B is
still consistent with the $\mu=0$ inflection point in Scheme A.  As
can be seen from fig.~\ref{fig:polyakov-fine-schemeAB}, the transition
in Scheme A happens at a considerably lower temperature than in Scheme
B, which may be taken as an indication of the width of the crossover
and associated uncertainty in the transition temperature.  It would be
useful to compare these scheme-dependent results with other quantities
sensitive to deconfinement, such as the static quark potential
\cite{Astrakhantsev:2018uzd} or the entropy of a static quark
\cite{Weber:2016fgn}.

\begin{table*}[tb]
\begin{tabular}{l|ccccc}
    $\mu a$ & 0.0 & 0.2 & 0.3 & 0.4 & 0.5 \\\hline
$aT_d^A(0.03)$ & & 0.1187\err{30}{45} & 0.072--0.125 & 0.057--0.113 & 0.055--0.094 \\
$aT_d^A(0.02)$ & & --- & 0.06--0.16 & 0.050--0.094 & 0.063--0.087 \\
$aT_d^A (j=0)$ & 0.1362\err{21}{27} & & 0.10(5) & 0.073(30) & 0.075(20) \\
$T_d^A$ (MeV) & 194(4) & $^*170(7)$ & 143(70) & 105(45) & 107(30)  \\ \hline
$aT_d^B(0.03)$ & & 0.1813\err{16}{27} & 0.1613\err{14}{9} & 0.1561\err{35}{60} & 0.1385\err{14}{8} \\
$aT_d^B(0.02)$ & & --- & 0.1565\err{3}{4} & 0.1529\err{5}{5} & 0.1423\err{13}{20} \\ 
$aT_d^B (j=0)$ & 0.164\err{7}{2} &     & 0.147(3) & 0.147(12) & 0.150(6) \\
$T_d^B$ (MeV) & 234(10) & $^*259(4)$  & 210(4) & 210(17) & 214(9) \\\hline
\end{tabular}
\caption{Inflection points $T_d(j)$ for $\braket{L(T)}$ at
  $ja=0.03 ,0.02$ using renormalisation schemes A (upper rows) and B
  (lower rows).  The uncertainties in the inflection points at $\mu=0$
  and in scheme B at all $\mu$ are purely statistical.  In scheme A,
  with the exception of $\mu a=0.2, ja=0.03$, it was only possible to
  determine a region where the transition occurs.  Also shown are
  estimates for the $j\to0$ extrapolated $T_d$ in each scheme.  For
  $\mu a=0.2$ the $ja=0.03$ values have been used.}
  \label{tab:Td}
\end{table*}
\begin{figure}[tb]
\includegraphics*[width=\colw]{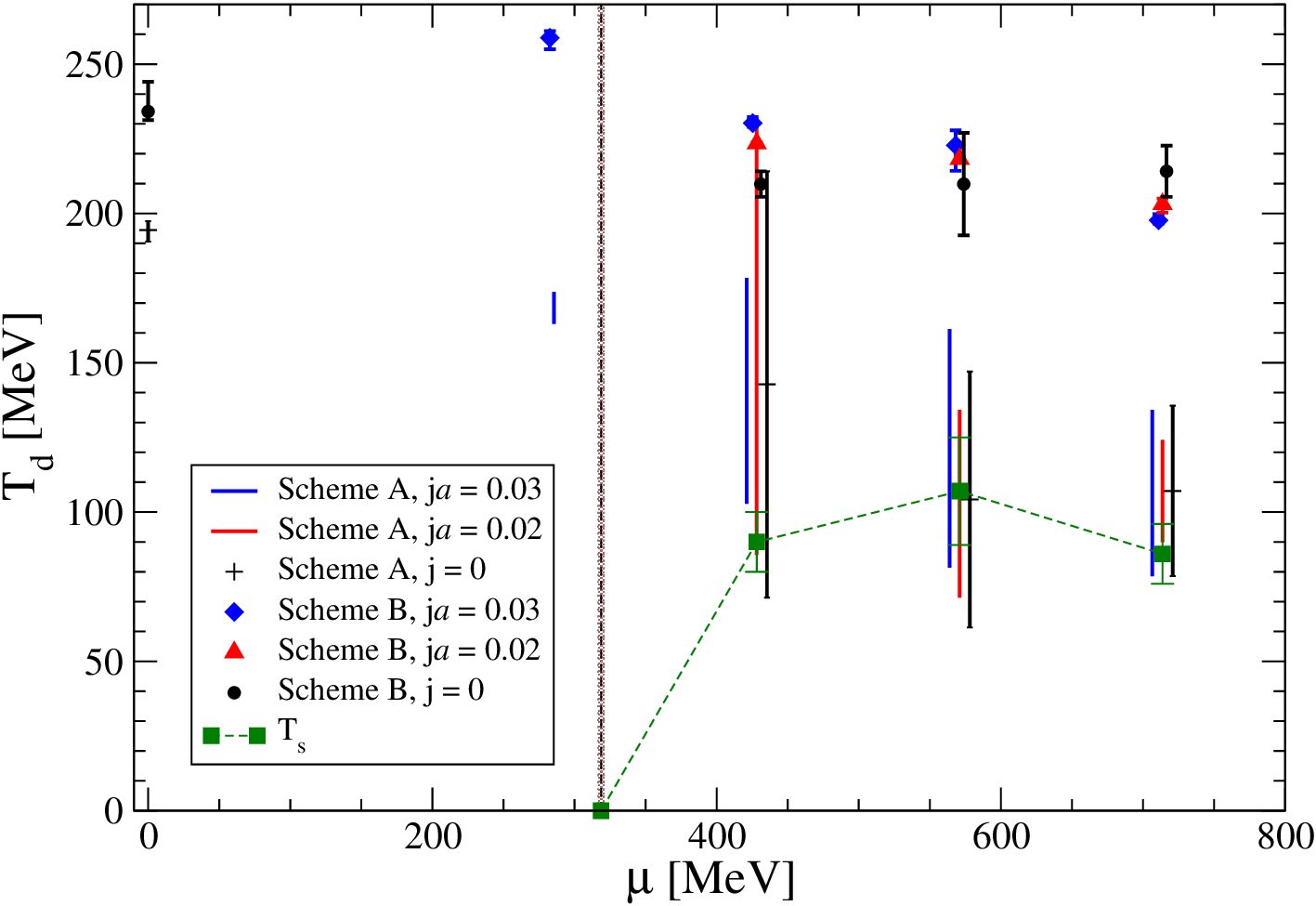}
\caption{The deconfinement temperature $T_d$ determined from the
  inflection point of the Polyakov loop, see text and
  Table~\ref{tab:Td} for details.  The vertical line denotes the onset
  chemical potential $\mu_o$.  Also shown is the superfluid transition
temperature $T_s$ from Table~\ref{tab:Ts}.}
\label{fig:deconfine}
\end{figure}
Our results for the deconfinement transition temperature are shown in
Table~\ref{tab:Td} and Fig.~\ref{fig:deconfine}.  We find a large
discrepancy between the two
schemes, and this discrepancy increases with $\mu$; however, the
uncertainties are large, in particular in scheme A which has a lower
$T_d$.  There are indications that, rather than decreasing
monotonically towards 0 at large $\mu$, $T_d$ approaches a constant value
which is above (scheme B) or close to (scheme A) the superfluid
transition temperature $T_s\approx90\,$MeV.  If this is
confirmed, it means there is no deconfining transition at $T=0$, and
that the superfluid phase remains confined at all $\mu$.

\begin{figure}[tb]
\includegraphics*[width=\colw]{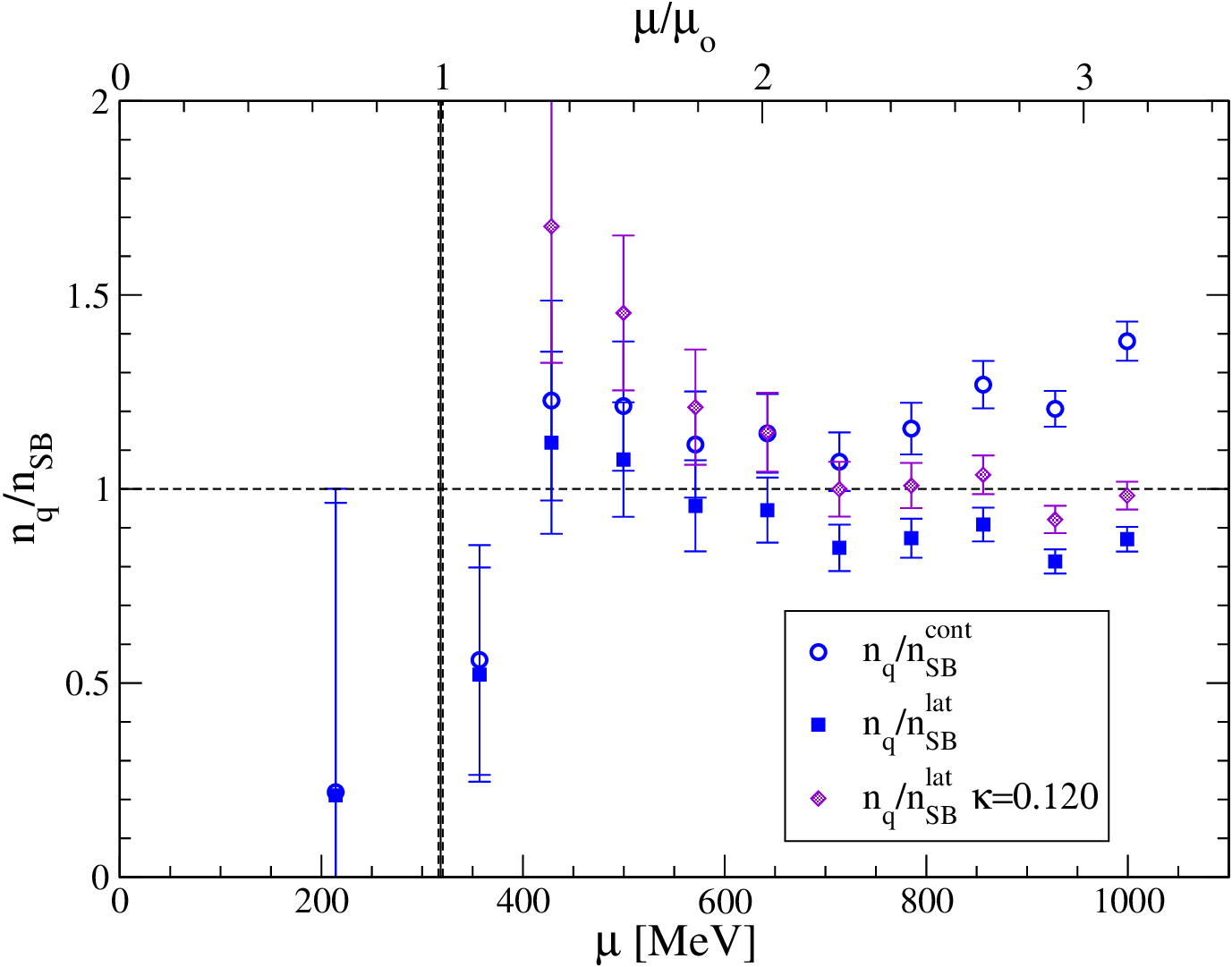}
\caption{The quark number density $n_q/n_{SB}$ from the fine ensemble
  at the lowest temperature ($T=43\,$MeV) extrapolated to $j=0$, with
  normalisation as described in the text.  The vertical lines indicate
  the location of the onset transition at $T=0$.}
\label{fig:density-fine-Nt32}
\end{figure}

\subsection{Quark number density}
\label{sec:nq-fine}

The final quantity to consider is the quark number density $n_q$,
which forms the basis for computing bulk thermodynamic properties such
as the pressure.  In figure~\ref{fig:density-fine-Nt32}
we show the quark number density extrapolated to zero diquark source
(using a linear extrapolation) at the lowest temperature studied,
corresponding to $N_\tau=32$.  As was the
case for the diquark condensate, we do not find significant evidence of a
deviation from a linear behaviour for the range of $j$-values we
have.  The data are plotted in dimensionless form by 
normalising by the density $n_{SB}$ for a gas of massless noninteracting
fermions.  Fig.~\ref{fig:density-fine-Nt32} shows results obtained
using two choices for $n_{SB}$: firstly using the continuum result
\begin{equation}
n_q^{SB} = \frac{N_fN_c}{3}\Big(\mu T^2 + \frac{\mu^3}{\pi^2}\Big)\,,
\label{eq:nq-cont}
\end{equation}
and secondly, using a form $n_{SB}^{\latt}$ evaluated as a mode
sum using the action (\ref{eq:Slatt},\ref{eq:Mwils}) with
$\kappa=0.125$ and $U_\mu\equiv1$ on a finite lattice as described in
\cite{Hands:2006ve}. Both forms were explored in \cite{Cotter:2012mb},
with the conclusion that while $n_{SB}^{\latt}$ offers a better
correction for UV artifacts at large $\mu a$, it is prone to
significant IR artifacts for $\mu\gtrsim\mu_o$, so that
$n_{SB}^{\cont}$ is preferred in this regime. In the current
work we have addressed this issue by evaluating $n_{SB}^{\latt}$
on a much larger spatial volume than that used for simulation,
specifically $96^3\times N_\tau$ (with $128^3\times N_\tau$ checked to
see that this was sufficient).  This removes the IR artifacts but
retains the UV corrections.

We find that $n_q/n_{SB}$ rises rapidly above the onset transition,
then descends to reach a plateau for $\mu\gtrsim500\,$MeV. This is in
qualitative agreement with the predictions of $\chi$PT shown in Fig.~3
of \cite{Hands:2006ve}. Within the fairly large errors resulting from
the $j\to0$ extrapolation the results obtained in the neighbourhood of
onset with continuum and lattice free fermions are compatible, but for
$\mu\gtrsim700\,$MeV the $n_{SB}^{\cont}$ curve continues to rise while
the lattice normalisation yields a plateau with $n_q/n_{SB}\lesssim1$;
for the reasons given in the previous paragraph this is the
normalisation we prefer and will use henceforth. Also shown in
Fig.~\ref{fig:density-fine-Nt32} is the ratio evaluated using free
massive Wilson fermions with $\kappa=0.120$; in this case the value of
$n_q/n_{SB}$ on the plateau is consistent with unity. Due to the
difficulties inherent in assigning a bare quark mass for interacting
Wilson fermions, we therefore do not draw any physical conclusions from the
plateau height at this stage.

An important observation is that there is no longer evidence for a
regime at high $\mu$ where $n_q/n_{SB}$ increases above its value on
the plateau (Cf.  Fig.~11 of \cite{Cotter:2012mb}; were the high-$\mu$
behaviour in that plot physical, then a corresponding rise would be
expected on the fine lattice at $\mu\approx700\,$MeV).  In conclusion, at
the lowest temperature studied the equation of state looks
``quarkyonic'', ie. with $n_q(\mu)\approx n_{SB}$, all the way along
the $\mu$-axis, with no evidence of a qualitative change associated
with deconfinement.  This is consistent with our results for the
Polyakov loop, which show no sign of a deconfinement transition at low
temperature, and the high-$\mu$ increase seen in \cphases\ is
therefore most likely a lattice artefact.

\begin{figure}[tb]
\includegraphics*[width=\colw]{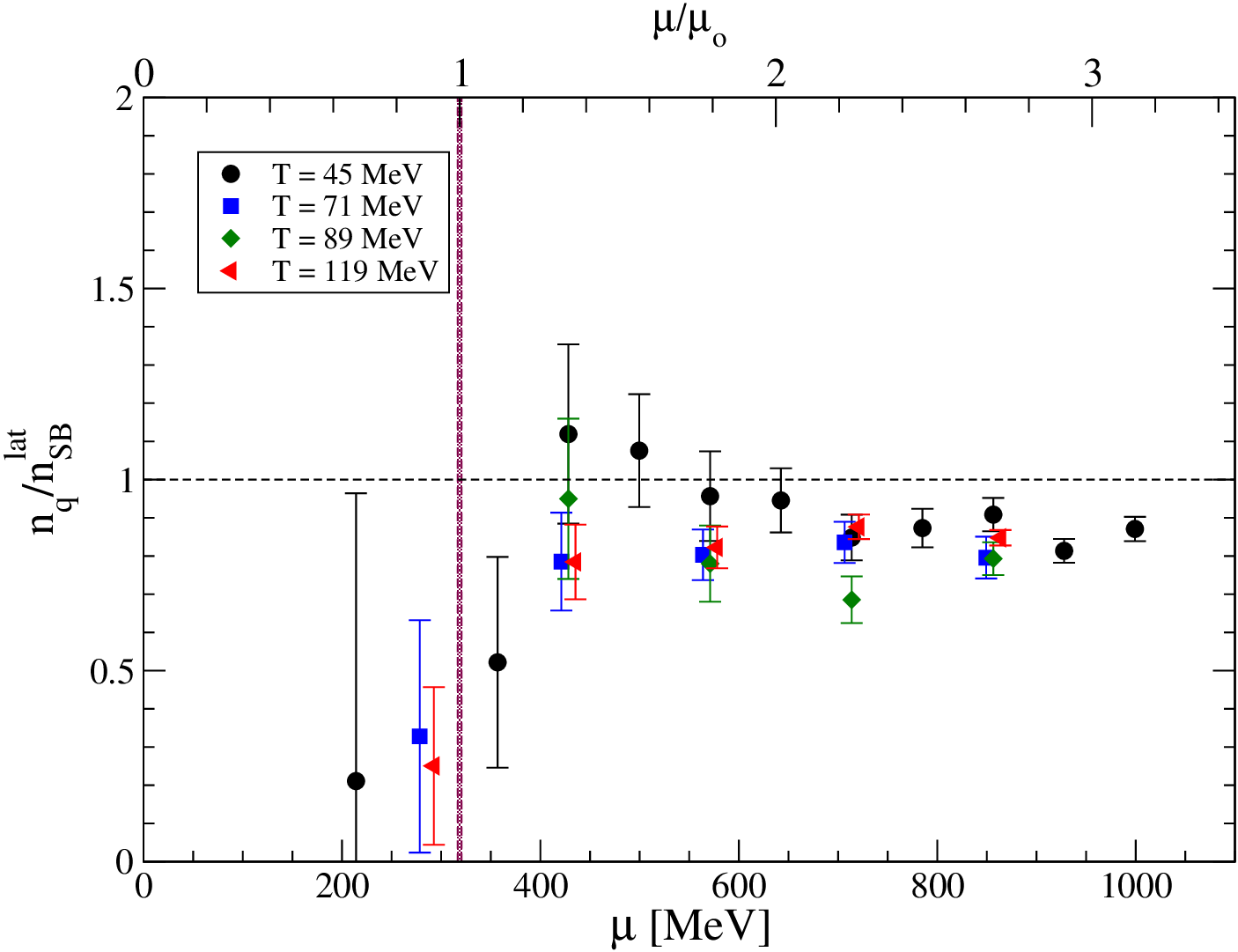}
\caption{The quark number density $n_q/n_{SB}$ from the fine
ensemble, extrapolated to $j=0$, for the four temperatures studied.} 
\label{fig:density-fine}
\end{figure}

As shown in Fig.~\ref{fig:density-fine}, this behaviour persists for
the lowest three temperatures, which according to the results in
section~\ref{sec:diquark-fine} are all in the superfluid region (or
near the transition temperature, for $N_\tau=16$).  The data suggest
the plateau value of $n_q/n_{SB}$ falls with increasing $T$, although
uncertainties following the $j\to0$ extrapolation are significant.  At
the highest temperature, $T=119\,$MeV ($N_\tau=12$), which falls in
the quark--gluon plasma phase, we see indications of $n_q/n_{SB}$
monotonically increasing with $\mu$, in qualitative agreement with the
findings of \cphases.

\begin{figure}[tb]
\includegraphics*[width=\colw]{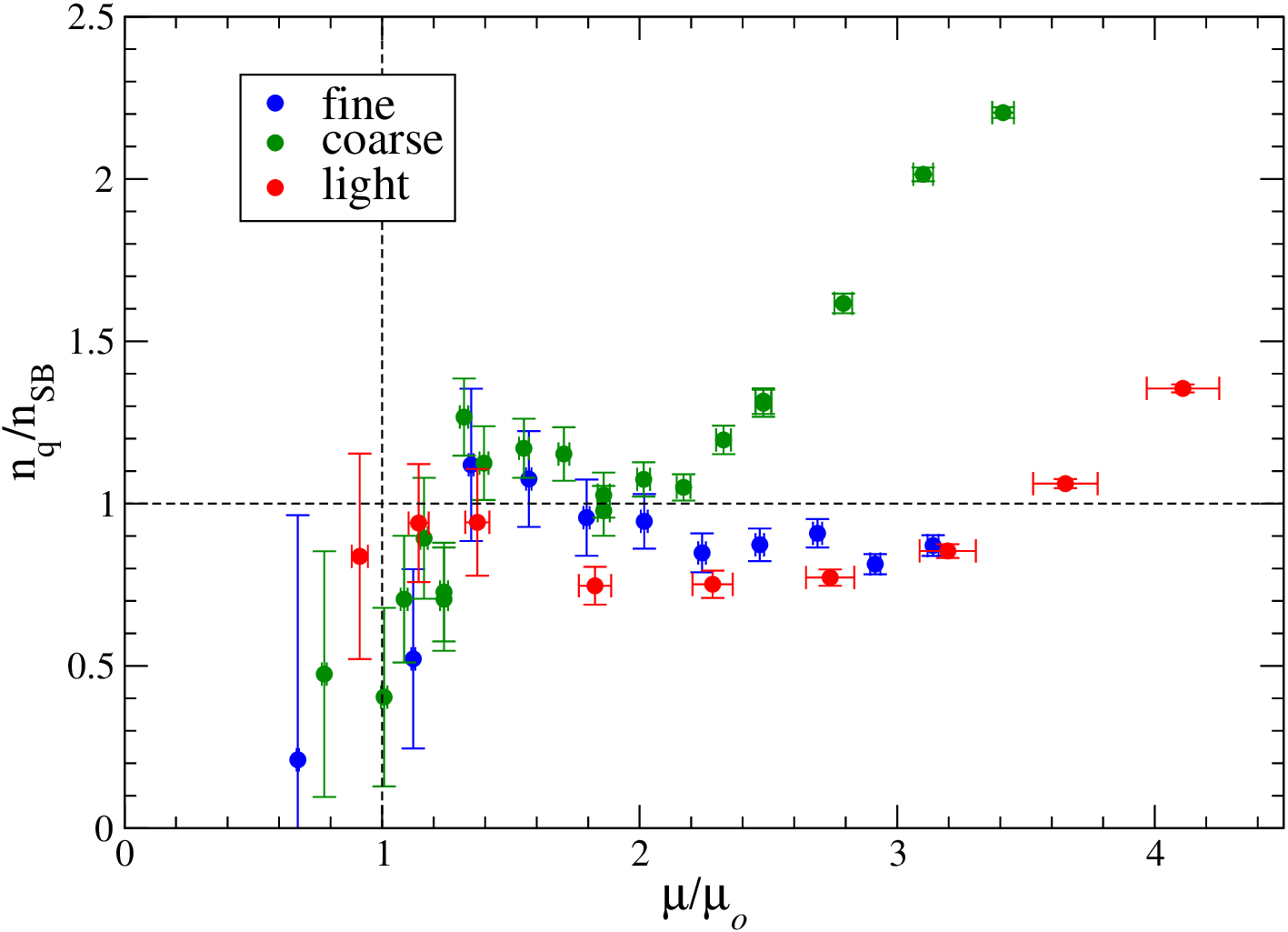}\\
\includegraphics*[width=\colw]{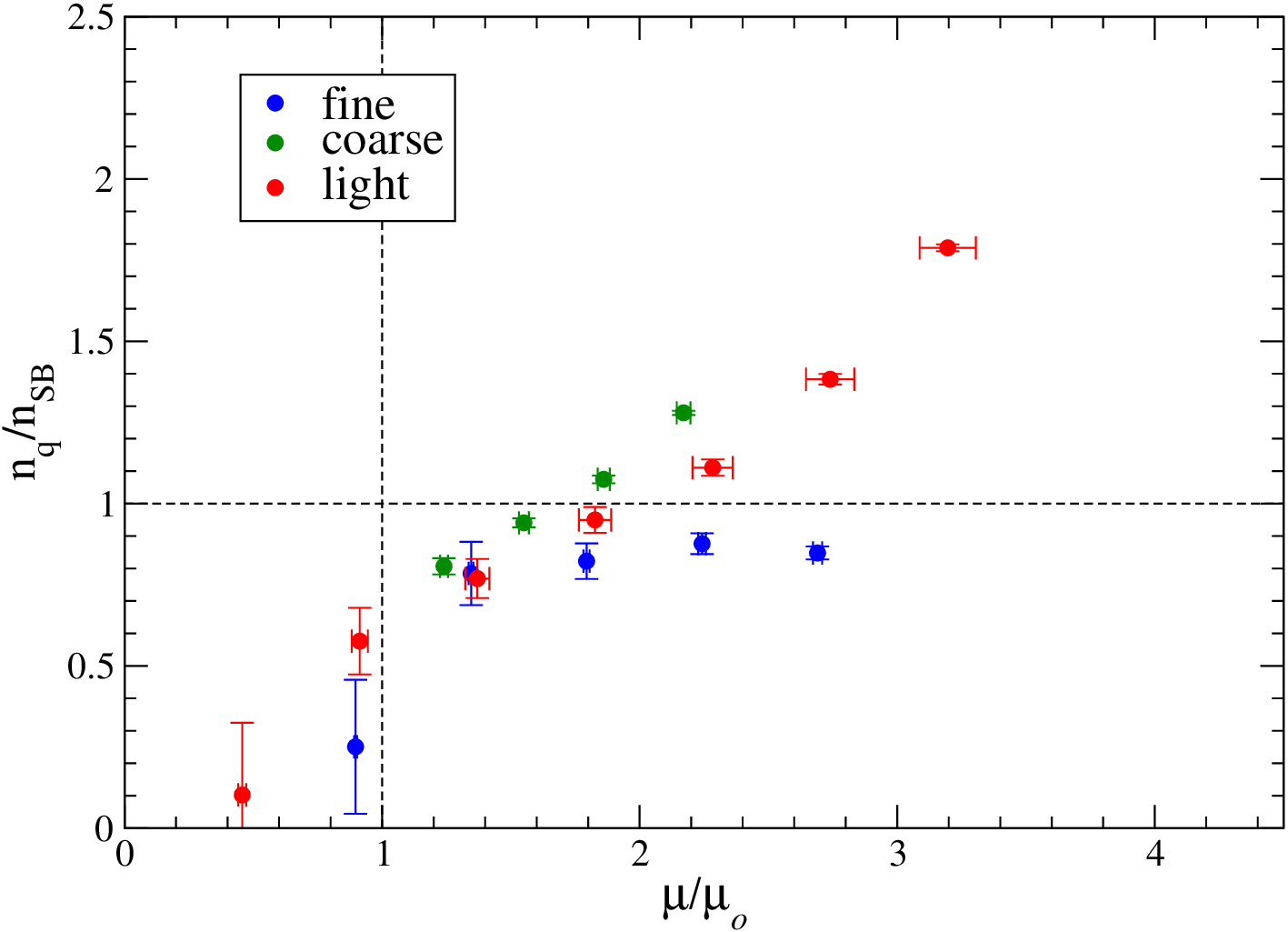}
\caption{The quark number density from the fine, coarse and light ensembles,
  divided by density in the noninteracting limit, as a function of
  chemical potential in units of the onset chemical potential
  $\mu_o$.  Upper panel: low temperature ($T\approx45\,$MeV); lower
  panel: high temperature ($T\approx130\,$MeV).}
\label{fig:density-coarse-fine-light}
\end{figure}

To make this comparison quantitative, we show the results from the
fine and coarse ensemble together, in units of the onset chemical
potential $\mu_o=m_\pi/2$ (allowing also for a comparison of results
for the different quark masses), in figure~\ref{fig:density-coarse-fine-light}.  At
low temperature (upper panel) there is quantitative agreement between
the two ensembles for $\mu\lesssim2\mu_o$.  For larger $\mu$ the rise
in $n_q/n_{SB}$ seen for the coarse ensemble (which might have
signalled a transition to a different state of matter) is absent for
the fine ensemble, and instead we see that $n_q$ remains close to
$n_{SB}$ throughout.  This is
consistent with our results for the Polyakov loop, which show no sign
of a deconfinement transition at low temperature, and the high-$\mu$
increase seen in \cphases\ is therefore most likely a lattice
artefact.  
The same pattern is repeated at high temperature (lower panel), where
now we see $n_q$ approach $n_{SB}$ from below.  However, part of the
difference between the two ensembles may in this case be due to the 
slightly different temperatures (120 vs 130 MeV for the fine and
coarse ensembles respectively).

\begin{figure}[tb]
\includegraphics*[width=\colw]{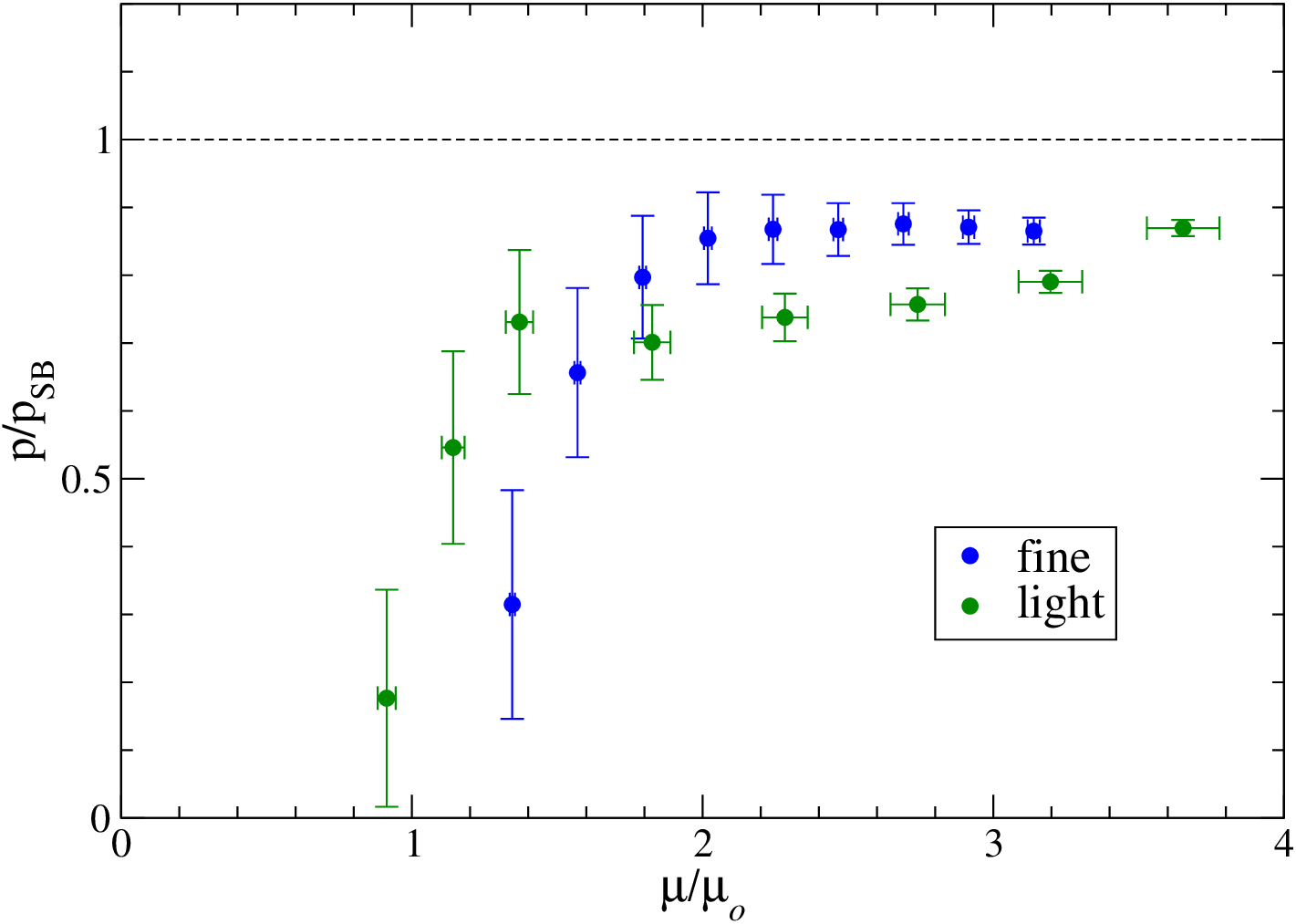}
\caption{The pressure $p/p_{SB}$ at low temperature
  ($T\approx45\,$MeV) from both fine and light ensembles,
  extrapolated to $j=0$ and plotted as a function of $\mu/\mu_o$.}
\label{fig:pressure}
\end{figure}

Next we discuss the pressure, which as outlined in \cite{Hands:2006ve,
Cotter:2012mb} can be obtained in the limit of low $T$ via  an integral of the
form $p=\int^\mu n_q(\mu^\prime) d\mu^\prime$. In order to arrive at the
dimensionless ratio $p/p_{SB}$ starting from our data there are
several different quadrature schemes available: here we use
\begin{equation}
\frac{p}{p_{SB}}
 = \frac{1}{p_{SB}^{\rm cont}}
 \int_{\mu_0}^\mu\frac{n_{SB}^{\cont}(\mu^\prime)}{n_{SB}^{\latt}(\mu^\prime)} n_q(\mu^\prime)d\mu^\prime,
\end{equation}
introduced as ``scheme II'' in \cite{Cotter:2012mb} ($\mu_0$ is the
lowest available value in the dataset) . Reassuringly, with
$n_{SB}^{\latt}$ now defined on a large spatial volume to eliminate IR
artifacts we find results compatible with those computed using scheme
I, which was not the case in \cite{Cotter:2012mb}. The results for
data extrapolated to $j\to0$, at low temperatures where there are
enough $\mu$-points to control the numerical integration, are shown in
Fig.~\ref{fig:pressure}.

The main result is that $p/p_{SB}$ increases sharply after onset,
reaching a plateau at $\mu/\mu_o\approx2$. It appears to approach the
plateau from below, in contrast to the $\chi$PT prediction that the SB
limit is approached from above~\cite{Hands:2006ve}.  While, as in the
discussion of $n_q/n_{SB}$, it is premature to assign a precise value
to the height of the plateau, it is worth recalling that e.g.\ in the
Van der Waals equation of state the ideal gas pressure receives a
downward correction due to \emph{attractive} forces between
particles.

\section{Results from light ensemble}
\label{sec:results-light}

\begin{figure}[tb]
\includegraphics*[width=\colw]{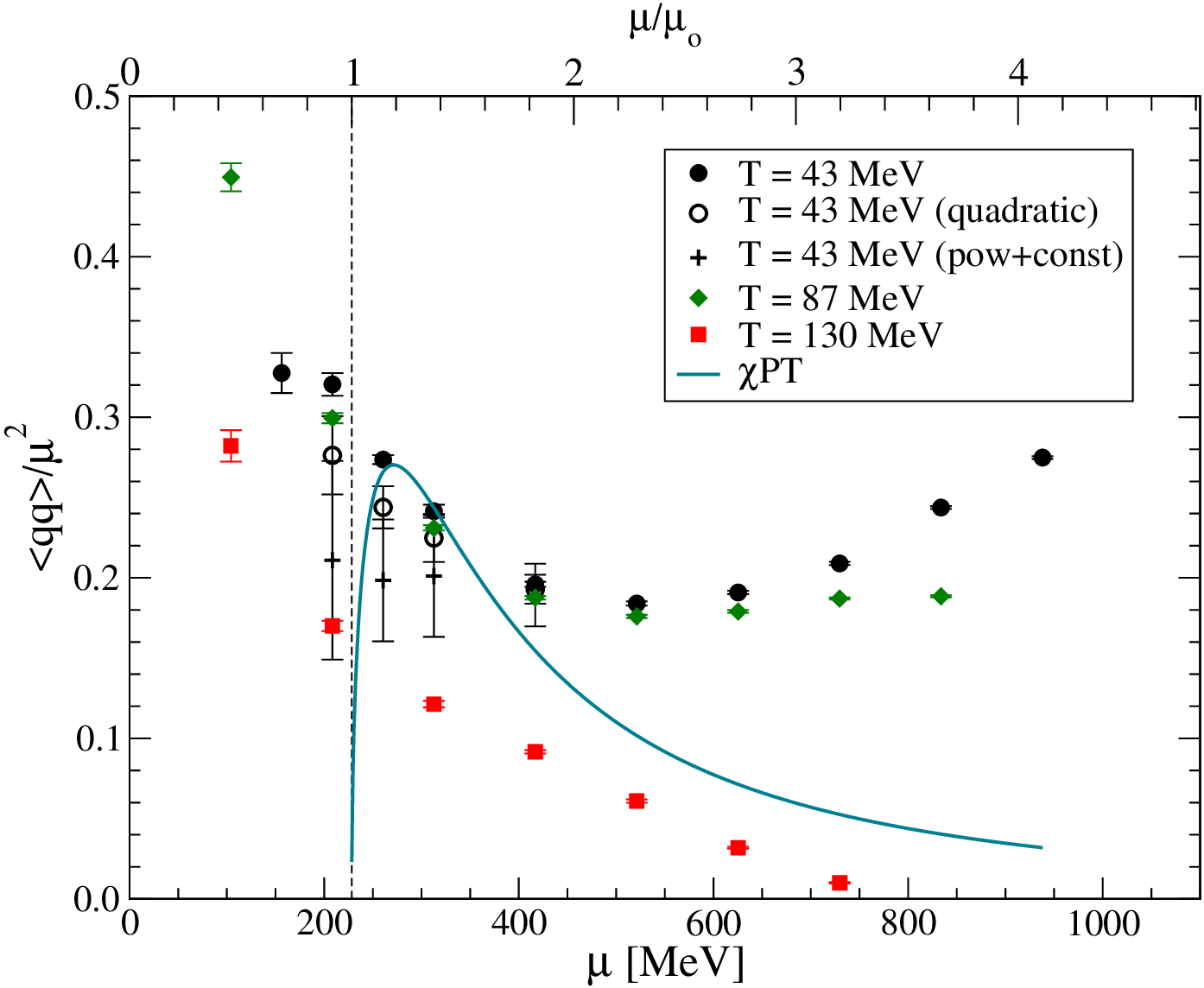}
\caption{The diquark condensate $\bra qq\ket/\mu^2$
from the light ensemble, extrapolated to $j=0$ for
$T=43,87,130$ MeV. For $T=43\,\mathrm{MeV}\; (N_\tau=24)$, three different extrapolation
forms have been used: linear $\qq=A+Bj$; quadratic $\qq=A+Bj+Cj^2$; and
constant + power $\qq=A+Bj^\alpha$. The vertical dashed line denotes the onset
transition, $\mu_o=m_\pi/2$.  The solid curve shows the prediction
\eqref{eq:qq-chiPT} from $\chi$PT, with an arbitrary
prefactor.}
\label{fig:qq-light}
\end{figure}

We now study the effect that reducing the quark mass may have on the
phase structure and equation of state.  While our parameters are still
very far from the chiral limit, this may give us an idea of which, if
any, qualitative changes may occur as we approach this limit.

Figure~\ref{fig:qq-light} shows the diquark condensate from the light
ensemble for our three different temperatures, extrapolated to $j=0$.
On the face of it, these results are qualitatively different from our
results with heavier quarks, in that the diquark condensate no longer
scales like $\mu^2$ in the region just above the onset transition.
This might be taken as an indication that a BEC window is opening up
where, rather than following the BCS scaling
$\braket{qq}\propto\mu^2$, the condensate behaves according to the
predictions \eqref{eq:qq-chiPT} of zero-temperature chiral perturbation theory.

To facilitate this comparison, we have included in
fig.~\ref{fig:qq-light} the $\chi$PT curve \eqref{eq:qq-chiPT}, with
an arbitrary prefactor, and we see that the data for $\mu_o<\mu<400\,$MeV
make contact with this curve.

This interpretation is, however, complicated by our lack of control of
the $j\to0$ extrapolation.  For nearly all points, we only have two
values of the diquark source ($ja=0.02, 0.04$) available and have used
a simple linear extrapolation.  For $\mu a=0.2, 0.25$ and 0.3 on the
$12^3\times24$ lattice we also have data for $ja=0.03$, and have also
used a quadratic extrapolation as well as a power law + constant form,
$\braket{qq}(j)=A+Bj^\alpha$.  While the quadratic extrapolation gives
results roughly consistent with the linear form, the power law form
gives a result which is consistent with BCS scaling at all $\mu$.
It should be noted that near the onset transition, we expect a
power-law scaling with $\alpha=1/3$.

\begin{figure}[tb]
\includegraphics*[width=\colw]{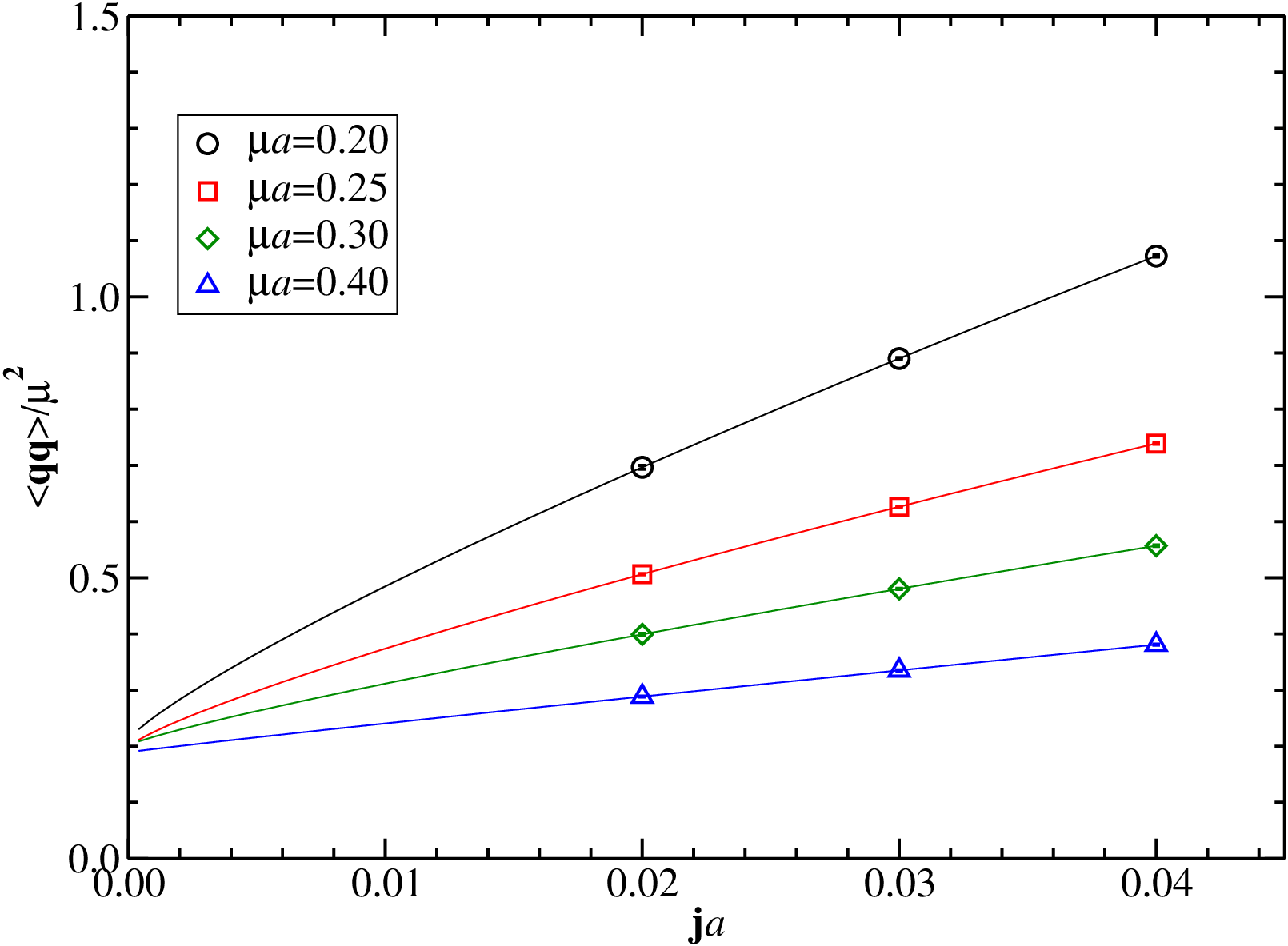}
\caption{The diquark condensate $\bra qq\ket$ for the lowest
  temperature on the light ensemble, as a function of the diquark
  source, for different values of the chemical potential $\mu$.  The
  solid lines are fits to a power-law + constant form $\qq=A+Bj^\alpha$.}
\label{fig:qq-light-jextrap}
\end{figure}

We also find that $\qq\neq0$ also for $\mu<\mu_o$, and we take this to
be an indication that the linear extrapolation breaks down in this
regime.  Figure~\ref{fig:qq-light-jextrap} shows the diquark
condensate as a function of $j$ for those chemical potentials where we
have 3 $j$-values at our disposal.  We see that the data are
consistent with a linear behaviour as was also found for the fine
ensemble, but may also be described with a power-law + constant form. 
A better control over the diquark source extrapolation, for
example along the lines of \cite{Brandt:2017oyy}, is required to
determine whether there is indeed a BEC window for these parameters.

The $\qq$ results for $T=87\,\mathrm{MeV}\; (N_\tau=12)$ are almost identical
to those for $T=43\,$MeV, except for $\mu\gtrsim600\,$MeV.  However, at
the highest temperature, $T=130\,$MeV ($N_\tau=8$) we see that the
diquark condensate is much smaller, and vanishes at high $\mu$.  It is
worth noting, however, that $\qq$ does not vanish at
all $\mu$, as was the case at comparable temperatures for the coarse
and fine ensembles.  This may be taken as a first indication of a
$\mu$-dependent superfluid to normal transition temperature.  A
controlled $j\to0$ extrapolation combined with temperature scans at
fixed $\mu$ would be required to draw a firm conclusion.

\begin{figure}[tb]
\includegraphics*[width=\colw]{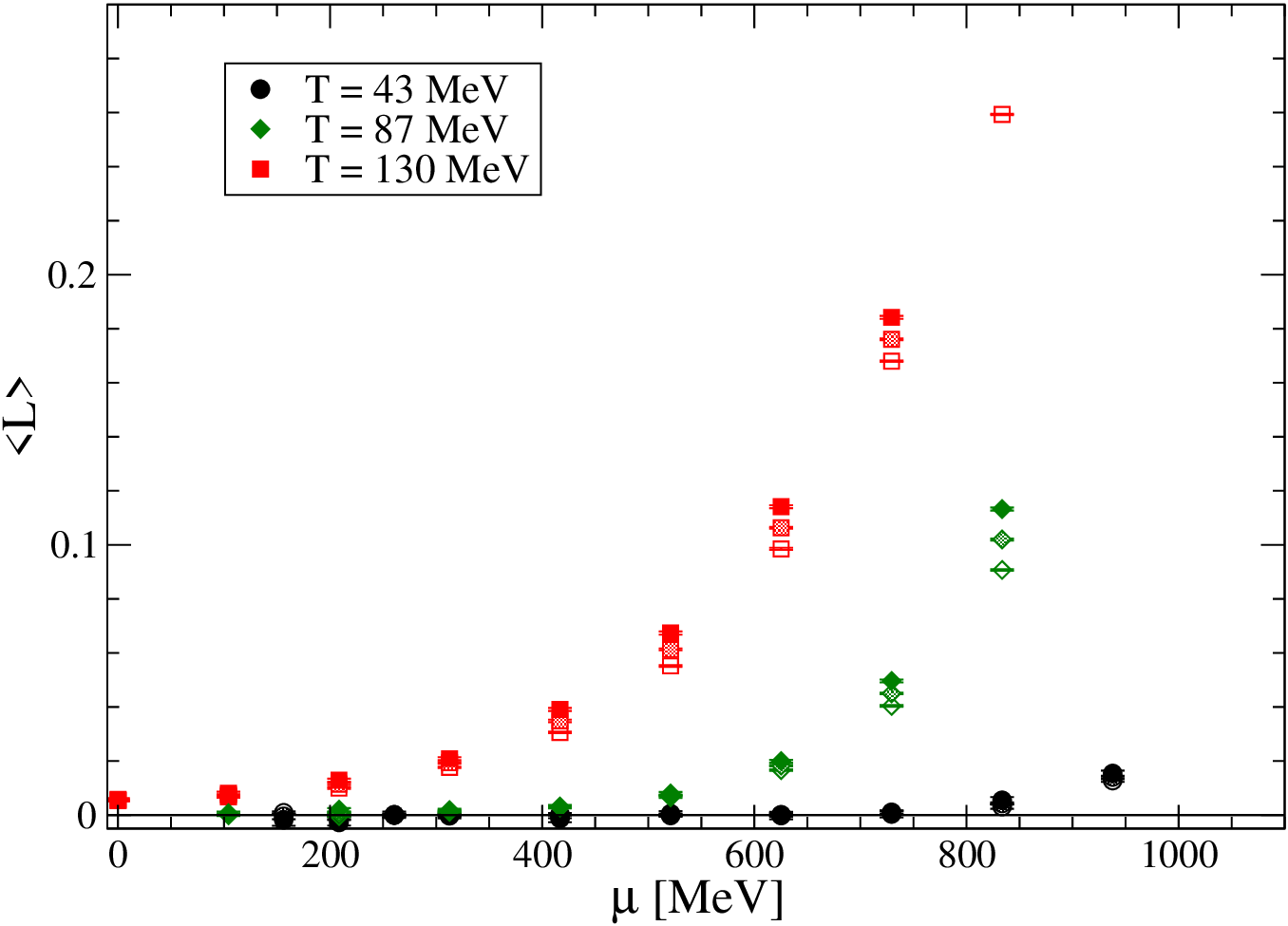}
\caption{The unrenormalised Polyakov loop from the light ensemble as a
  function of chemical potential $\mu$, for all temperatures.  The open
  symbols are for $ja=0.04$, the shaded symbols for $ja=0.02$, and the
  filled symbols are extrapolated to $j=0$. 
}
\label{fig:polyakov-light-mu}
\end{figure}
Looking now at the deconfinement transition,
figure~\ref{fig:polyakov-light-mu} shows the Polyakov loop as function
of chemical potential for our three temperatures.  In the absence of a
$\mu=0$ temperature scan to fix the renormalisation constant, we show
the unrenormalised Polyakov loop.  This does not affect the shape of
each of the curves, only the relative magnitude of the data at
different temperatures.  We see the same picture as before, where 
$\braket{L}\simeq0$ at low temperature (except for
$\mu\gtrsim800\,$MeV, corresponding to $\mu a\gtrsim0.8$, which in the
light of previous results we attribute to a 
lattice artefact), but increasing with $\mu$ at higher temperature.
At $T=130\,$MeV ($N_\tau=8$), we note that $\braket{L}\neq0$ also at
$\mu=0$, suggesting that this temperature is near the transition
region for this ensemble.

\begin{figure}[tb]
\includegraphics*[width=\colw]{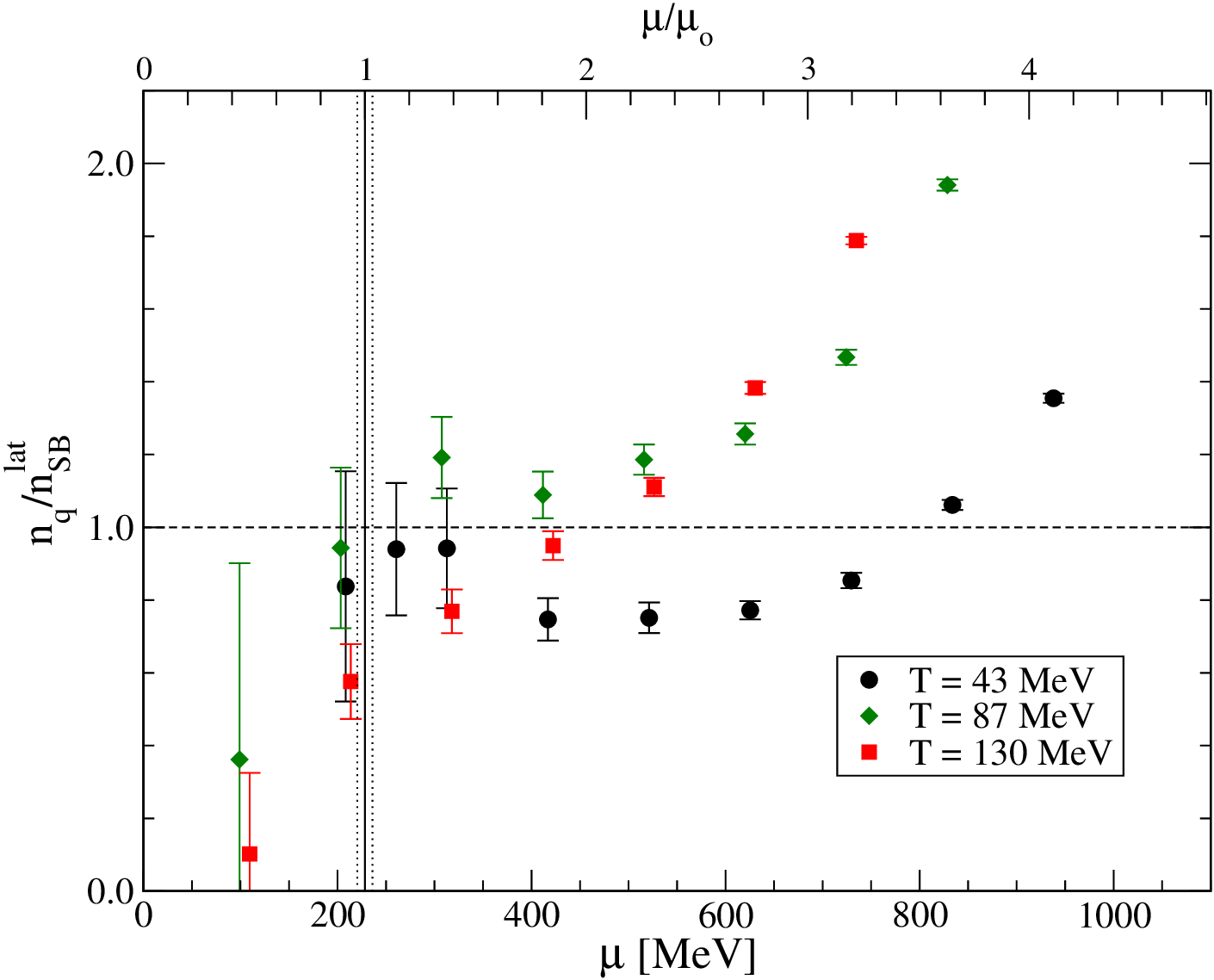}\\
\includegraphics*[width=\colw]{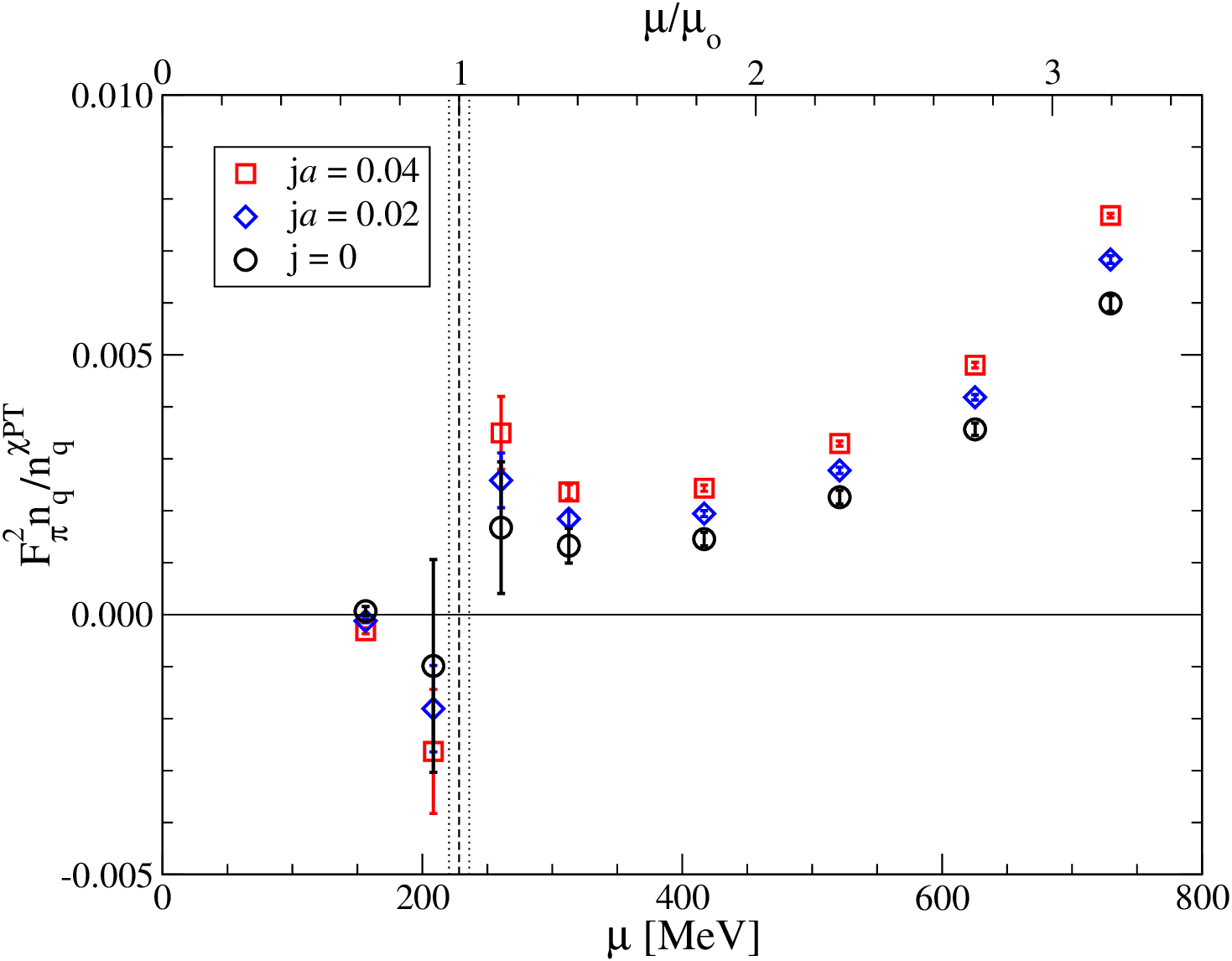}
\caption{Quark number density $n_q$ from the light ensemble at $j=0$
  divided by $n_{SB}^{\latt}$ (top) and at $T=43\,$MeV divided by the
  chiral perturbation theory form $\mu(1-\mu_o^4/\mu^4)$ (bottom).
  The dashed vertical line indicates the onset transition at zero
  temperature.}
\label{fig:density-light}
\end{figure}
In fig.~\ref{fig:density-light} we show the quark number density
$n_q/n_{SB}$ calculated using the same procedure outlined in
Sec.~\ref{sec:nq-fine}. At low temperature, this has a plateau for
$\mu>\mu_o$, in line with our previous results, while at high
temperature we see a steeper increase with $\mu$, as 
seen before, which is much more pronounced than that for the fine lattice
shown in Fig.~\ref{fig:density-fine}. On a coarse lattice the impact of UV
artefacts in this regime cannot be excluded. 
However, as can be seen in the lower panel of
fig.~\ref{fig:density-light}, the quark number density at the lowest
temperature is also consistent with the $\chi$PT prediction
\eqref{eq:nq-chiPT}, as indicated by the $j\to0$ extrapolated curve
tending to a non-zero constant as $\mu\to\mu_{o+}$.
It therefore remains an open question whether we for these parameters
have a window characterised by a Bose--Einstein condensate of tightly
bound diquarks.

In figure~\ref{fig:density-coarse-fine-light} we compare our results for the
quark number density with those obtained on the coarse ensemble in
\cphases.  We see
that the smaller quark mass has a large quantitative effect, giving a
much lower number density.  Because $\mu_o a$ is now smaller,
the increase in $n/n_{SB}$ at large $\mu$, which is likely to be a
lattice artefact, now only appears at larger $\mu/\mu_o$.  At high
temperature, however, we do not see any plateau in $n_q/n_{SB}$, as
was the case for the fine ensemble, but instead a monotonic increase
with $\mu$.

Finally, we note that results for $p/p_{SB}$ for the light ensemble are included
in Fig.~\ref{fig:pressure}. The notable feature is that the rise from zero
following onset is now much steeper, and an approximate plateau established by
$\mu/\mu_o\lesssim1.5$. As before, we are reluctant to over-interpret the
disparity in plateau height between fine and light ensembles.

\section{Conclusions and outlook}
\label{sec:conclude}

We have performed lattice simulations of two-colour QCD at different
lattice spacings and different quark masses, as a step towards
the continuum and chiral (or light-quark) limits of this theory.  Our
investigation of the phase structure of the theory in the $(T,\mu)$
plane has yielded results in qualitative agreement with earlier
studies \cprev; most notably it has confirmed the quarkyonic phase at
low $T$ and intermediate to high $\mu$.  In this phase, quarks are
confined but the bulk thermodynamics as well as the diquark condensate
behave as if the system consists of a Fermi sphere of weakly
interacting quarks.  This phase is found to extend to larger $\mu$ as
the lattice spacing or the quark mass is reduced.  A striking result
is that at the lowest temperatures explored the quark number density
is found to be very close to the Stefan--Boltzmann value.
Because we have not yet determined the physical quark mass, it is not
possible to state with confidence whether the ratio $n_q/n_{SB}$ is
greater or less than one in this regime. However,
Fig.~\ref{fig:density-light} does suggest that the ratio falls as the
quark mass decreases.

The superfluid phase transition temperature is found to be independent
of $\mu$ at least for our heavier quark mass, corresponding to
$m_\pi/m_\rho=0.8$.  The transition temperature $T_s\approx90\,$MeV is in
quantitative agreement with that found on a coarser lattice in
\ctrans.   This qualitative behaviour also
agrees with what has been found for QCD at high isospin density
\cite{Brandt:2017oyy}.  For lighter quarks, however, there are
indications that $T_s$ may increase with $\mu$. 

Although we find a deconfinement transition at high temperature, the
deconfinement transition previously seen at low $T$ and high $\mu$
appears to be a lattice artefact --- at $T\sim45\,$MeV the Polyakov
loop appears to increase from zero at $\mu a\sim0.8$ for all lattice
parameters.  Simulations using staggered fermions at a relatively high
temperature $T>100\,$MeV
\cite{Braguta:2016cpw,Astrakhantsev:2018uzd} have found a
deconfinement transition at $\mu\approx750\,$MeV characterised by the
string tension dropping to zero.  This is not in contradiction with
the absence of any deconfinement transition at $T\approx45$\,MeV.

The location of the high-temperature deconfinement transition is not
yet clear.  Our results suggest a broad crossover with a transition
temperature $T_d$ that decreases with increasing $\mu$.  The analysis
is complicated by lack of precise data for the renormalised Polyakov
loop at low temperature, as well as a small but nontrivial diquark
source dependence.  There are indications that $T_d$ may tend to a
constant at large $\mu$ and that $T_d\gtrsim T_s$ for all $\mu$.

There remain significant uncertainties relating to the extrapolation
to zero diquark source $j$, with our data unable to distinguish
between a linear $j$-dependence and the nonanalytic behaviour expected
at least in the vicinity of the onset transition.  This may be
mitigated by adopting the reweighting method introduced in
\cite{Brandt:2017oyy} and implemented in the context of QC$_2$D in
\cite{Iida:2019rah}; this will be left for future investigations.

Another source of uncertainty is the use of relatively small volumes,
with $N_s<N_\tau$ for our lowest temperatures.  Additional simulations
on larger volumes would be required to reliably determine if our
assignment of a nonzero temperature to these ensembles is correct.  We
note that finite volume effects were studied previously in
Ref.~\cphases; there it was found that the results for the various
physical quantities, and in particular the Polyakov loop, agreed
between the two volumes considered, suggesting that finite volume
effects are indeed small.

A reliable continuum extrapolation would require significantly finer
lattices than those used in this study.  The first step towards
further reducing the lattice spacing errors would be to employ an
improved fermion action.  This, together with algorithmic
improvements, would also allow us to study lighter quark masses which
are out of bounds with the unimproved Wilson action used here.

\begin{acknowledgments}
This work used the DiRAC Blue Gene Q Shared Petaflop system at the University
of Edinburgh, operated by the Edinburgh Parallel Computing Centre on behalf of
the STFC DiRAC HPC Facility (www.dirac.ac.uk). This equipment was funded by BIS
National E-infrastructure capital grant ST/K000411/1, STFC capital grant
ST/H008845/1, and STFC DiRAC Operations grants ST/K005804/1 and ST/K005790/1.
DiRAC is part of the National E-Infrastructure.
JIS and TSB
acknowledge the support of Science Foundation Ireland grants
11-RFP.1-PHY3193 and 11-RFP.1-PHY3193-STTF-1.
SJH was supported by STFC grant ST/L000369/1.
We acknowledge the support of COST Action CA15213 ``Theory of
relativistic heavy-ion collisions''.
PG thanks the Institute for Theoretical Physics of the
University of M\"unster where part of the work was done.
JIS expresses his deep appreciation for the hospitality of the Galileo
Galilei Institute, Florence, where this work was 
completed.
\end{acknowledgments}

\bibliography{density}

\end{document}